\documentclass[aps,prc,twocolumn,floatfix,showpacs,preprintnumbers,amsmath,amssymb,nofootinbib,groupedaddress]{revtex4-1}
\usepackage{color}
\usepackage{graphicx}
\usepackage{dcolumn}    % Align table columns on decimal point
\usepackage{multirow}
\usepackage{bm}         % bold math
\usepackage{sidecap}
\usepackage{hyperref}   % use for hypertext links, including those to external documents and URLs
\usepackage{amsmath}
\usepackage{graphicx}
\usepackage{float}
\usepackage{adjustbox}
\usepackage{caption}
\usepackage{subcaption}

\usepackage{mathtools} % $ \prescript{14}{2}{\mathbf{C}} $
\usepackage[version=3]{mhchem} % for chemical equations: \ce{^{227}_{90}Th+}

\usepackage[dvipsnames,usenames]{xcolor}
\usepackage{mathrsfs,natbib}
\usepackage{epsf,amssymb,amsbsy,amsfonts,amssymb,amsmath}
\usepackage{slashed}
\usepackage{comment}

\usepackage{hyperref}
\usepackage[justification=raggedright]{caption}
\definecolor{dark-red}{rgb}{0.,0.,0}
\definecolor{dark-blue}{rgb}{0.,0.,1}
\definecolor{medium-blue}{rgb}{0,0,1}
\hypersetup{
    colorlinks, linkcolor={dark-red},
    citecolor={dark-blue}, urlcolor={medium-blue}
}

\begin{document}
% >>>>>>>>>>>>>>>>>>>>>>>>>>>>>>>>>>>>>>>>>>>>>>>>>>>>>>>>>>>>>>>>>>>>
% TITLE AND AUTHORS.
%
\title{Ground State Properties of Charmed Hypernuclei with Mean Field Approach}
%\title{Investigation of Ground State Properties of Charmed nuclei with Mean Field Approach}
%\title{Charmed nuclei with Mean Field Approach}

\author{H. G\"{u}ven}
%\affiliation{Institut de Physique Nucl\'eaire, Universit\'e Paris-Sud, IN2P3-CNRS, F-91406 Orsay Cedex, France}
\affiliation{Universit\'e Paris-Saclay, CNRS/IN2P3, IJCLab, 91405 Orsay, France}
\affiliation{Physics Department, Yildiz Technical University, 34220 Esenler, Istanbul, Turkey}

\author{K. Bozkurt}
\affiliation{Physics Department, Yildiz Technical University, 34220 Esenler, Istanbul, Turkey}
\affiliation{Universit\'e Paris-Saclay, CNRS/IN2P3, IJCLab, 91405 Orsay, France}

\author{E. Khan}
\affiliation{Universit\'e Paris-Saclay, CNRS/IN2P3, IJCLab, 91405 Orsay, France}

\author{J. Margueron}
%\author[0000-0001-8743-3092]{J\'er\^ome Margueron}
\affiliation{Univ Lyon, Univ Claude Bernard Lyon 1, CNRS/IN2P3, IP2I Lyon, UMR 5822, F-69622, Villeurbanne, France}
\date{\today}

% >>>>>>>>>>>>>>>>>>>>>>>>>>>>>>>>>>>>>>>>>>>>>>>>>>>>>>>>>>>>>>>>>>>>
% ABSTRACT, PACS.
%
\begin{abstract}
Closed shell charmed hypernuclei
\ce{^5_{$\Lambda_c$}Li}, \ce{^17_{$\Lambda_c$}F}, \ce{^41_{$\Lambda_c$}Sc}, \ce{^57_{$\Lambda_c$}Cu}, \ce{^133_{$\Lambda_c$}Sb} and \ce{^209_{$\Lambda_c$}Bi}
are calculated within Hartree-Fock approach by using three different force sets derived from microscopic Brueckner-Hartree-Fock calculations of $\Lambda$ hypernuclei.
Ground state properties (binding energies, $\Lambda_c$ separation energies, $\Lambda_c$ single particle energies and $\Lambda_c$ densities) of charmed nuclei are examined. %JM Compatible results with those obtained by different models from the literature are found.
Due to the Coulomb repulsion between protons and the $\Lambda_c$ baryon, charmed hypernuclei are most bound for $16\leq$A$\leq 41$, where \ce{^17_{$\Lambda_c$}F} can be considered as an excellent candidate to measure charmed hypernuclei.
The competition between the attractive nucleon-$\Lambda_c$ interaction and the Coulomb repulsion is discussed, and we compare $\Lambda$ and $\Lambda_c$ hypernuclei properties.
\end{abstract}

\maketitle
\section{Introduction}

The discovery of charmed hadron $\Lambda_c$~\cite{Aubert1974,Augustin1974,Cazzoli1975,Goldhaber1976,Knapp1976,Peruzzi1976,Peruzzi1976}, after the one of the (strange) $\Lambda$ hyperon~\cite{Danysz1953}, opened the possibility to produce charmed hypernuclei, an extension of very exotic hypernuclei.
From a theory viewpoint, such systems were first discussed in the seminal work of Dover and Kahana in 1977~\cite{Dover1977}, where bound charmed nuclei were predicted, based on an interaction potential generated by SU(4) symmetry.
Meantime, experimental efforts to investigate charmed nuclei were performed in Dubna in the 1970's and 1980's~\cite{Batusov1976,Batusov1981,Lyukov1989}.
In these experiments, only three candidate were found, namely \ce{_${\Lambda_c}$ Li}, \ce{_${\Lambda_c}$ B} and \ce{_${\Lambda_c}$ N},
where the separation energy of $\Lambda_c$ was measured to be between 0 and 10 MeV~\cite{Lyukov1989}.
In the future, GSI-FAIR (Gesellschaft für Schwerionenforschung–Facility for Antiproton and Ion Research) and JPARC (Japan Proton Accelerator Research Complex)
facilities are expected to produce sufficient charmed particles to generate more charmed hypernuclei~\cite{Riedl2007,Shyam2017,Krein2018}.

The key ingredient in the theoretical description of charmed hypernuclei is
%The key point to investigate charmed hypernuclei lies in the description of
the nucleon-$\Lambda_c$ (N$\Lambda_c$) interaction.
Historically, SU(4) flavor symmetry was considered in order to determine N$\Lambda_c$ potential, similarly to
%as an analogy to
the phenomenological nucleon-nucleon (NN) and N$\Lambda$ potentials.
With such potentials, the ground state properties were obtained using few-body methods~\cite{Bhamathi1981,Bando1982,Bando1983,Gibson1983,Bando1985}.
However this method reaches its limits as the nuclear mass number grows, and it becomes more relevant to consider density functional theory (DFT) approaches.
%method is more reliable from few-body calculation.
%Accordingly this topic has been revisited in
In the past few years, charmed hypernuclei have been revisited, based on DFT, e.g.
% by using more sophisticated methods such as
relativistic mean field (RMF) model~\cite{Tan2004,Tan2004v2}, quark meson coupling (QMC) model~\cite{Tsushima2003,Tsushima2003v2,Tsushima2004,Wu2020}, or the Brueckner-Hartree-Fock
%state of art perturbative many body
approach~\cite{Vidana2019}.
The results obtained from these DFT approaches differ, especially because of their different N$\Lambda_c$ interaction.
%are slightly different from each other, due to their various N$\Lambda_c$ interactions.
For instance, the coupling constants between charmed baryons and mesons are usually obtained from the quark counting rules in QMC and RMF models.
Another approach for describing N$\Lambda_c$ potential, is using lattice QCD simulation:
the central part and the tensor part of N$\Lambda_c$ interaction are calculated by HAL QCD Collaboration~\cite{Miyamoto2018}.
It is found that  $\Lambda_c$ hypernuclei can exist between A$=12$ to A$\approx 50$~\cite{Miyamoto2018}.
%\JM{In which paper? Give reference}.
Then, Haidenbauer and Krein extrapolated the N$\Lambda_c$ interaction from the HAL QCD one, with physical pion mass and chiral effective field theory~\cite{Haidenbauer2018}.
Vidaña et al.  also calculated charmed hypernuclei, using SU(4) extension of the meson-exchange hyperon-nucleon potential of the Jülich group, where phase shifts of Model B and Model C are compatible with the low energy region of extrapolation from Haidenbauer and Krein~\cite{Vidana2019}.
Recently, charmed hypernuclei from \ce{^5_{$\Lambda_c$}Li} to \ce{^{209}_{$\Lambda_c$}Bi} have been calculated with the same approach developed by Haidenbauer et al.~\cite{Haidenbauer2020}.
On the contrary to HAL QCD results ~\cite{Miyamoto2018}, both Vidaña et al. and Haidenbauer et al. (see Refs.~\cite{Haidenbauer2020,Vidana2019} for details) predicted bound $\Lambda_c$ hypernuclei for the A$\geq 50$ region, where $\Lambda_c$ binding energies are compatible with previous QMC calculations~\cite{Tsushima2003,Tsushima2003v2,Tsushima2004}.
The $\Lambda_c$ bound states of QMC calculations are inside the gap defined between model A and Model C of Vidaña et al. for the A$\leq 50$ region.
However, the interactions calculated from the HAL QCD with the chiral effective field theory~\cite{Haidenbauer2020} results in a less attractive N$\Lambda_c$ potential compared to 
the interactions derived from hyperon-nucleon potential of the Jülich group~\cite{Vidana2019},
%where the binding energy difference is about $7$($48$) MeV compared to Model C (Model A) for \ce{^{209}_{$\Lambda_c$}Bi} charmed hypernucleus.
where the binding energy differences are $7.2\pm1.5$ MeV for Model C and $49\pm1.5$ MeV for Model A for \ce{^{209}_{$\Lambda_c$}Bi} charmed hypernucleus.
%Despite these similarities they obtained different predictions for heavy charmed hypernuclei ~\cite{Miyamoto2018}. In the A$\geq 50$ region, their results\JM{Which results?} are compatible with other theoretical calculations~\cite{Tsushima2003,Tsushima2003v2,Tsushima2004,Wu2020}.
Another recent calculation was done by Wu et al., with QMC model, where coupling constants are obtained from both the naive quark counting rules and the HAL QCD predictions~\cite{Wu2020}.
They confirm the results of Vidaña et al.~\cite{Vidana2019} for the coupling constant obtained from the naive quark counting rules,
where bound $\Lambda_c$ states are calculated up to \ce{^{209}_{$\Lambda_c$}Pb}.
However,
they also find compatible result with HAL QCD Collaboration~\cite{Miyamoto2018} from the coupling constant obtained from HAL QCD simulation,
where bound $\Lambda_c$ states are only possible for the A$\leq 52$ region.

In this work, we explore the ground state properties of charmed hypernuclei with non-relativistic DFT approach, where the N$\Lambda_c$ interaction is fixed within the DFT framework.
%Well known N$\Lambda$ potentials.
In the Skyrme-Hartree-Fock framework, the N$\Lambda$ channel is based on a G-matrix calculation starting from various bare interactions:
NSC89, NSC97a and NSC97f (Nijmegen Soft Core Potentials)~\cite{Schulze2013}.
The density functional, deduced from the G-matrix calculation in uniform matter, will be generically named as
DF-NSC89, DF-NSC97a and DF-NSC97f, hereafter, see Ref.~\cite{Guven2018} for more details.
%\JM{Give ref to our previous work where we introduced this notation}
The oldest DF-NSC89 functional can reproduce with a good accuracy
the experimental single particle energies of $\Lambda$ hyperon for light hypernuclei such as \ce{^{5}_{$\Lambda$}He} or \ce{^{13}_{$\Lambda$}C}, but for the heavier hypernuclei like \ce{^{41}_{$\Lambda$}Ca} or \ce{^{209}_{$\Lambda$}Pb}, DF-NSC97a and DF-NSC97f give  results close to the experimental data~\cite{Schulze2013,Vidana2001}.
In order to generate N$\Lambda_c$ interaction, we use the  approach introduced by Starkov et al., where the similarity of the in-medium interaction between N$\Lambda_c$ and N$\Lambda$ channels is considered~\cite{Starkov1986}.
We detail this scheme in the next section.

It should be noted that the notation of charmed nuclei is changing in the literature, whether the charge of $\Lambda^+_c$ is considered or not. In this work, we consider the chemical convention, where the element name refers to the total charge of the nucleus (including protons and $\Lambda^+_c$ charges), as well as the mass is the total mass counting nucleons and hyperons).
In the following, we will investigate \ce{^5_{$\Lambda_c$}Li} which can be decomposed as [$^4$He$+\Lambda_c$],
as well as  \ce{^{17}_{$\Lambda_c$}F} ([$^{16}$O$+\Lambda_c$]),
 \ce{^{41}_{$\Lambda_c$}Sc} ([$^{40}$Ca$+\Lambda_c$]),
 \ce{^{57}_{$\Lambda_c$}Cu} ([$^{56}$Ni$+\Lambda_c$]),
 \ce{^{133}_{$\Lambda_c$}Sb} ([$^{132}$Sn$+\Lambda_c$]),
 \ce{^{209}_{$\Lambda_c$}Bi} ([$^{208}$Pb$+\Lambda_c$]).
Such nuclei have been selected in the present study since the nucleon cores are all closed shell, which allows to neglect deformation and pairing effects.
They are expected to be almost spherical, and the
%Starting from lightest charmed hypernuclei $^5$(He$+\Lambda_c$), we calculate, $^{17}$(O$+\Lambda_c$), $^{41}$(Ca$+\Lambda_c$), $^{57}$(Ni$+\Lambda_c$), $^{133}$(Sn$+\Lambda_c$) and
%$^{209}$(Pb$+\Lambda_c$) where all core nuclei are closed shell and spherical in order to neglect deformation and pairing effects.
possible small deformation coming from $\Lambda_c$ is neglected because of the doubly neutron and proton closed shells.
We therefore consider Equal Filling Approximation (EFA) for the occupation probability of the 1s $\Lambda_c$-states~\cite{Martin2008}.
Finally, we shall investigate binding energies, single particle spectra, Coulomb repulsion and density distribution of $\Lambda_c$, in related charmed hypernuclei.

The paper is organized in the following way:
Sec.~\ref{TF}, details the Skyrme-Hartree-Fock framework and the calculation of N$\Lambda_c$ interaction.
In Sec.~\ref{Results}, the ground state properties of charmed hypernuclei are discussed.
Finally, a brief summary and some concluding remarks are given in Sec.~\ref{Conclusions}.

%
%
%In this work, ground state properties of single and multi $\Lambda$ hypernuclei are investigated with Hartree-Fock-Bogoliubov (HFB) formalism considering $\Lambda$ pairing interactions.
%On this purpose we neglect the $\Lambda$ spin-orbit interaction
%which is estimated to be very small~\cite{hashimoto2006,motoba2008}.
%The three body interactions such as NN$\Lambda$~\cite{18,Lonardoni2013,Lonardoni2015} is effectively included from the functional approach.
%We have added zero range pairing force to the NN and $\Lambda\Lambda$ channels, opening the possibility to calculate accurately open-shell nuclei.
%
%The HFB equations for multi-strange hypernuclei are presented in Sec. II.
%The general features of shell evolution for multi strange hypernuclei are discussed in Sec. III.
%The possibility of N$\Lambda$ pairing is discussed in Sec. IV, and, in Section 5 results with and without pairing interaction are discussed.
%Conclusions and outlooks are given in the last Sec.VI.

\section{Theoretical framework}\label{TF}
Considering a non-relativistic system, composed of interacting nucleons $N$ and $\Lambda_c's$, the total Hamiltonian reads,
\begin{equation}\label{e1}
  \widehat{H}=\widehat{T}_{N}+\widehat{T}_{\Lambda_c}+\widehat{H}_{N N}+\widehat{H}_{N \Lambda_c},
\end{equation}
where $\widehat{T}_N$($\widehat{T}_{\Lambda_c}$) are the kinetic energy operators for nucleons($\Lambda_c's$) and $\widehat{H}_{N N}$($\widehat{H}_{N \Lambda_c}$) are the interaction operator terms acting between $N$ and $N$($N$ and $\Lambda_c$) species.

%\begin{equation}\label{e1}
%  \widehat{H}=\widehat{T}_{A}+\widehat{T}_{B}+\widehat{H}_{A A}+\widehat{H}_{A B},
%\end{equation}
%where $\widehat{T}_A$ are the kinetic energy operators and $\widehat{H}_{A B}$ are the interaction operator terms acting between $A$ and $B$ species ($A,B=N,\Lambda_c$)\JM{Why not using directly $A=N$ and $B=\Lambda$?}.

\subsection{Mean-field approximation}

In the mean field approximation, the ground state of the system is the tensor product $|\Psi_N\rangle \otimes |\Psi_\Lambda\rangle$, where $ |\Psi_N\rangle$ ($|\Psi_\Lambda\rangle$) is a slater determinant of the nucleon ($\Lambda_c$) states.
The total Hamiltonian~(\ref{e1}) can be turned into a density functional $\epsilon(\rho_N,\rho_{\Lambda_c})$, function of the particle densities $\rho_N$ and $\rho_{\Lambda_c}$, as
$\widehat{H}=\int \epsilon(\rho_N,\rho_{\Lambda_c}) d^3r$. The energy functional $\epsilon$ is often expressed as~\cite{Cugnon2000,Vidana2001},
\begin{eqnarray}\label{e3}
\epsilon(\rho_N,\rho_{\Lambda_c})=\frac{\hbar}{2m_N}\tau_N+\frac{\hbar}{2m_{\Lambda_c}}\tau_{\Lambda_c} +\epsilon_{N N}(\rho_N)  \nonumber \\
+\epsilon_{N \Lambda_c}(\rho_N,\rho_{\Lambda_c})+\epsilon_\mathrm{Coul}(\rho_{p},\rho_{\Lambda_c}),
\end{eqnarray}
where $\tau_{N}$ ($\tau_{\Lambda_c}$) is the nucleonic (charmed hyperonic) kinetic energy density, $\epsilon_\mathrm{Coul}(\rho_{p},\rho_{\Lambda_c})$ is the Coulomb interaction which depends on proton and $\Lambda_c$ densities,
and $\epsilon_{N N}$ and $\epsilon_{N \Lambda_c}$
%$\epsilon_{i j}$
are the interaction terms of the energy density functional describing the NN and N$\Lambda_c$  channels.

In the following, the nucleonic terms will be deduced from the well known SLy5 Skyrme interaction~\cite{Bender2003}, widely used for the description of the structure of finite nuclei.
In order to calculate the $N\Lambda_c$ interaction channel, we use a scaling relation introduced by Starkov et. al.,~\cite{Starkov1986} and defined as,
%JM It is shown that the $N\Lambda_c$ channel can be obtained from the $N\Lambda$ interaction in a simple equation from:
\begin{equation}\label{e3v2}
\epsilon_{N \Lambda_c}=K \epsilon_{N \Lambda},
\end{equation}
where the $K$ factor is $0.8$, from the estimation of coupling constant of $\sigma$ and $\omega$ mesons.
Accordingly, the $N\Lambda$ interaction is given by a density functional $\epsilon_{N \Lambda}$, adjusted to BHF predictions in uniform matter~\cite{Cugnon2000,Vidana2001},
\begin{equation}\label{e5}
\epsilon_{N \Lambda}(\rho_N,\rho_\Lambda)=-f_1(\rho_N)\rho_N\rho_\Lambda+ f_2(\rho_N)\rho_N\rho_\Lambda^{5/3}.
\end{equation}
In the $N\Lambda$ channel, the spin-orbit doublets are experimentally undistinguishable, typically around $100$-$200$ KeV~\cite{Hashimoto2006,Motoba2008}, and the spin-orbit interaction among $\Lambda$ particles can safely be neglected~\cite{Khan2015}.
An interesting mechanism, based on the quark sub-structure of hadrons explains the strong reduction of the spin-orbit in the $N\Lambda$ channel~\cite{Chanfray2020}.
Even smaller spin-orbit splitting could be expected for $N\Lambda_c$ channel, due to the large mass of $\Lambda_c$ baryons ($m_{\Lambda_c}=2286.46$ MeV, see Ref.~\cite{Tanabashi2019} for details)~\cite{Krein2018,Chanfray2020}.
Therefore, we also omit spin-orbit interaction in the case of the $N\Lambda_c$ interaction.

We now recall the details of the $N\Lambda$ channel, where the following density functionals are considered: DF-NSC89~\cite{Cugnon2000}, DF-NSC97a~\cite{Vidana2001}, DF-NSC97f~\cite{Vidana2001}.
The functions $f_{1-2}$ in Eq.~(\ref{e5}) are expressed as,
\begin{eqnarray}
f_1(\rho_N)&=&\alpha_1-\alpha_2\rho_N+\alpha_3\rho_N^2, \label{e7} \\
f_2(\rho_N)&=&\alpha_4-\alpha_5\rho_N+\alpha_6\rho_N^2, \label{e8}
\end{eqnarray}
where $\alpha_{1-6}$ are constants given in Table \ref{t1}.

\begin{table}
\centering
\caption{Parameters of the functionals DF-NSC89, DF-NSC97a and DF-NSC97f.}
\label{t1}
\tabcolsep=0.20cm
\def\arraystretch{1.5}
\begin{tabular}{lllllll}
\hline\hline
Force           & $\alpha_1$ & $\alpha_2$ & $\alpha_3$ & $\alpha_4$ & $\alpha_5$ & $\alpha_6$  \\ \hline
DF-NSC89        & 327        & 1159       & 1163       & 335        & 1102       & 1660        \\
DF-NSC97a       & 423        & 1899       & 3795       & 577        & 4017       & 11061       \\
DF-NSC97f       & 384        & 1473       & 1933       & 635        & 1829       & 4100        \\
\hline\hline
\end{tabular}
\end{table}

%\begin{table}
%\centering
%\caption{Parameters of the modified functionals of DF-NSC89, DF-NSC97a and DF-NSC97f \cite{28}.}
%\label{t2}
%\tabcolsep=0.1cm
%\def\arraystretch{1.5}
%\begin{tabular}{llllllllll}
%\hline\hline
%Functional                                                     & $\alpha_1$ & $\alpha_2$ & $\alpha_3$ & $\alpha_4$ & $\alpha_5$ & $\alpha_6$ & $\alpha_7$& $\alpha_8$ & $\alpha_9$ \\ \hline
%\begin{tabular}[c]{@{}l@{}}DF-NSC89\\ +EmpC\end{tabular}  & 327        & 1159       & 1163       & 335        & 1102       & 1660       &     22.81     & 0          & 0          \\
%
%\begin{tabular}[c]{@{}l@{}}DF-NSC97a\\ +EmpC\end{tabular} & 423        & 1899       & 3795       & 577        & 4017       & 11061      &       21.12   & 0          & 0          \\
%
%\begin{tabular}[c]{@{}l@{}}DF-NSC97f\\ +EmpC\end{tabular} & 384        & 1473       & 1933       & 635        & 1829       & 4100       &       33.25   & 0          & 0          \\ \hline\hline
%
%\end{tabular}
%\end{table}

In uniform nuclear matter, the single particle energies read,
\begin{equation}\label{e10}
\epsilon_N(k)=\frac{\hbar^2 k^2}{2m_N^*}+v_N^\mathrm{matt} \hbox{ and }
\epsilon_\Lambda(k)=\frac{\hbar^2 k^2}{2m_\Lambda^*}+v_\Lambda^\mathrm{matt},
\end{equation}
where the potentials $v_N$ and $v_\Lambda$ derive from the energy functional as
\begin{eqnarray}\label{e12}
v_N^\mathrm{matt}(\rho_N,\rho_\Lambda)&=& v_N^\mathrm{Skyrme}+\frac{\partial \epsilon_{N \Lambda} }{\partial \rho_N}\,, \\
v_\Lambda^\mathrm{matt}(\rho_N,\rho_\Lambda)&=& \frac{\partial \epsilon_{\Lambda \Lambda} }{\partial \rho_\Lambda} + \frac{\partial \epsilon_{N \Lambda} }{\partial \rho_\Lambda}\,.
\end{eqnarray}
In the case of hypernuclei, the energy functional defined in Eq.~(\ref{e5}) is corrected by the effective mass term as  (see Ref.~\cite{Margueron2017} and therein),
\begin{eqnarray}
\epsilon_{NN}^\mathrm{nucl}\!&=&\!\epsilon_{NN}\!-\!\frac{3  \hbar^2}{10m_N}\rho_N^{5/3}\!\left(\frac{6\pi^2}{g_N}\right)^{2/3}\!\left[\frac{m_N}{m^*_N}-1\right]\!\!, \label{e16}\\
\epsilon_{N \Lambda}^\mathrm{nucl}\!&=&\!\epsilon_{N\Lambda}\!-\frac{3 \hbar^2}{10m_\Lambda}\rho_\Lambda^{5/3}\!\left(\frac{6\pi^2}{g_\Lambda}\right)^{2/3}\!
\left[\frac{m_\Lambda}{m^*_\Lambda}-1\right]\!,\label{e17}
\end{eqnarray}
where the effective mass correction for the N$\Lambda$ part can be expressed as a polynomial in the nucleonic density $\rho_N$ as~\cite{Cugnon2000},
\begin{equation}\label{e19}
  \frac{m_\Lambda^*(\rho_N)}{m_\Lambda}=\mu_1-\mu_2\rho_N+ \mu_3\rho_N^2-\mu_4\rho_N^3.
\end{equation}
The values for the parameters $\mu_{1-4}$ are given in Tab.~\ref{t3}.
\begin{table}
\centering
\caption{The parameters of the $\Lambda$-effective mass.}
\tabcolsep=0.45cm
\def\arraystretch{1.5}
\label{t3}
\begin{tabular}{lllll}
\hline\hline
Force     & $\mu_1$ & $\mu_2$ & $\mu_3$ & $\mu_4$ \\ \hline
DF-NSC89  & 1.00                         & 1.83                         & 5.33                         & 6.07                       \\
DF-NSC97a & 0.98                         & 1.72                         & 3.18                         & 0                            \\
DF-NSC97f & 0.93                         & 2.19                         & 3.89                         & 0                            \\ \hline\hline
\end{tabular}
\end{table}
The effective mass correction of nucleon is given from Skyrme interaction~\cite{Bartel2008}.

\subsection{Charmed hypernuclei}

We now generate the $\Lambda_c$ interaction for charmed hypernuclei, employing Eq.~(\ref{e3v2}), which transforms the N$\Lambda$ force into a N$\Lambda_c$ force:
\begin{equation}\label{e20}
 \epsilon_{N \Lambda_c}(\rho_N,\rho_{\Lambda_c})=-f_1(\rho_N)\rho_N\rho_{\Lambda_c}+ f_2(\rho_N)\rho_N\rho_{\Lambda_c}^{5/3},
\end{equation}
where the functions $f_{1-2}$ are
\begin{eqnarray}
f_1(\rho_N)&=&\alpha^c_1-\alpha^c_2\rho_N+\alpha^c_3\rho_N^2, \label{e21} \\
f_2(\rho_N)&=&\alpha^c_4-\alpha^c_5\rho_N+\alpha^c_6\rho_N^2, \label{e22}
\end{eqnarray}
where $\alpha^c_{1-6}$ are new constants for $\Lambda_c$.
These constants are given in Table \ref{t4}, corresponding to K=0.8 in Eq.~(\ref{e3v2}).
\begin{table}
\centering
\caption{Parameters of the functionals DF-NSC89-C, DF-NSC97a-C and DF-NSC97f-C.}
\label{t4}
\tabcolsep=0.15cm
\def\arraystretch{1.5}
\begin{tabular}{lllllll}
\hline\hline
Force           & $\alpha^c_1$ & $\alpha^c_2$ & $\alpha^c_3$ & $\alpha^c_4$ & $\alpha^c_5$ & $\alpha^c_6$  \\ \hline
DF-NSC89-C  & 261.6 & 927.2  & 930.4  & 268   & 881.6  & 1328    \\
DF-NSC97a-C & 338.4 & 1519.2 & 3036   & 461.6 & 3213.6 & 8848.8  \\
DF-NSC97f-C & 307.2 & 1178.4 & 1546.4 & 508   & 1463.2 & 3280    \\
\hline\hline
\end{tabular}
\end{table}
Hereafter we call these force sets as DF-NSC89-C, DF-NSC97a-C and DF-NSC97f-C.
However, since nothing is known about effective mass for $\Lambda_c$,
we use the same effective mass parameters than for the N$\Lambda$ interaction defined in Eq.~(\ref{e19}).
Finally the energy functional for N$\Lambda$ interaction becomes,
\begin{equation}
\epsilon_{N \Lambda_c}^\mathrm{nucl}\!=\!\epsilon_{N\Lambda_c}\!-\frac{3 \hbar^2}{10m_{\Lambda_c}}\rho_{\Lambda_c}^{5/3}\!\left(\frac{6\pi^2}{g_{\Lambda_c}}\right)^{2/3}\!
\left[\frac{m_\Lambda}{m^*_\Lambda}-1\right]\!.\label{e23}
\end{equation}

Since $\Lambda_c$ is a positively charged particle, one needs to include Coulomb interaction to the total energy functional defined in Eq.(\ref{e3}).
Coulomb interaction is decomposed in direct and exchange terms as,
\begin{equation}\label{e24}
\epsilon_\textrm{Coul}(\rho_{p},\rho_{\Lambda_c})=E_\textrm{Coul}^D+E_\textrm{Coul}^E .
\end{equation}
The direct Coulomb term is,%\JM{why $i\neq j$?}
\begin{equation}\label{e25}
E_\textrm{Coul}^D=\sum_{i\neq j}^{} \frac{e^2}{2}\int d^3 \mathbf{r} d^3 \mathbf{r^\prime} \rho_i(\mathbf{r})\frac{1}{\mid \mathbf{r}-\mathbf{r^\prime} \mid }\rho_j(\mathbf{r^\prime}),
\end{equation}
where $i,j=p,\Lambda_c$.
It should be noted that the $p \Lambda_c$ channel is repulsive with respect of direct Coulomb interaction.

Considering the Slater approximation, the exchange term reads,
\begin{equation}\label{e26}
E_\textrm{Coul}^E=-e^2 \frac{3}{4}\left ( \frac{3}{\pi} \right )^\frac{1}{3}\int d^3 \mathbf{r}(\rho_p^{4/3}+\rho_{\Lambda_c}^{4/3}).
\end{equation}
The exchange term is attractive for all charged particles.

\subsection{Hartree-Fock equations}

We are now ready to extend our calculation to generate ground state properties of charmed hypernuclei.
The spherically symmetric Hartree-Fock (HF) framework is considered for present single-charmed hypernuclei, close to doubly magic nuclei.
In the HF approach, the Schrödinger equation can be obtained by minimizing the total energy defined in Eq.~\eqref{e3},
and using Skyrme model for the nucleonic part~\cite{Bender2003},
The usual Schrödinger equation is then obtained $(i = N,\Lambda_c)$,
\begin{eqnarray}\label{e27}
 \left [ -\nabla.\frac{\hbar^2}{2m^*_i(r)}\nabla+V_i(r)-iW_i(r)(\nabla \times \sigma) \right ]\psi_{i,\alpha}(r) \nonumber \\
 =-e_{i,\alpha}\psi_{i,\alpha}(r), \enspace
\end{eqnarray}
where $V_i$ is the interaction potential and $W_i$ is the spin-orbit potential~\cite{Ring1980} which is naturally obtained by Skyrme model~\cite{Bender2003}.
Since the spin-orbit interaction is neglected for $\Lambda_c$ channel, $W_{\Lambda_c}=0$ in Eq.~\eqref{e27}.

The interaction potentials are extracted by taking the functional derivative of the energy with respect to the densities.
Following the DFT framework, the nucleon potential reads
\begin{eqnarray}\label{e28}
V_N(\textbf{r}) \equiv v_N^\mathrm{Skyrme}+\frac{\partial \epsilon_{N \Lambda_c} }{\partial \rho_N}+\frac{\partial}{\partial \rho_N} \bigg [ \frac{m_\Lambda}{m_\Lambda^* (\rho_N)}  \bigg] \nonumber \\
\times \bigg [ \frac{\tau_{\Lambda_c}}{2m_{\Lambda_c}}-\frac{3}{5} \frac{(3\pi^2)^{2/3}\hbar^2}{2m_{\Lambda_c}} \rho_{\Lambda_c}^{5/3} \bigg] \nonumber \\
+v_\textrm{Coul,p}^D(r)+v_\textrm{Coul,p}^E(r)\,,
\end{eqnarray}
and the $\Lambda_c$ potential is given by
\begin{eqnarray}\label{e29}
 V_{\Lambda_c}(\textbf{r})\equiv \frac{\partial \epsilon_{N{\Lambda_c}}}{\partial \rho_{\Lambda_c}}
-\bigg[\frac{m_\Lambda}{m_\Lambda^*(\rho_N)}-1 \bigg] \frac{(3\pi^2)^{2/3} \hbar^2}{2m_{\Lambda_c}}\rho_{\Lambda_c}^{2/3} \nonumber \\
+v_{\textrm{Coul},\Lambda_c}^D(r)+v_{\textrm{Coul},\Lambda_c}^E(r)\,, \enspace
\end{eqnarray}
where $v_{\textrm{Coul},i}^D(r)$ ($v_{\textrm{Coul},i}^E(r)$) is the direct (exchange) Coulomb potential, which can be extracted by taking derivatives for related density
$\rho_i$ ($i=\textrm{p},\Lambda_c$).
Direct Coulomb potential is
\begin{equation}\label{e30}
v_{\textrm{Coul},i}^D(r)=e^2\int d^3 \mathbf{r}^\prime\frac{1}{\left |\mathbf{r}-\mathbf{r}^\prime  \right |}\rho_{ch}(\mathbf{r}^\prime),
\end{equation}
where $\rho_{ch}=\rho_{p}+\rho_{\Lambda_c}$ is the charge density.
We consider the extension of the Slater approximation for multiple types of charged particles,
giving for the exchange Coulomb potential
\begin{equation}\label{e31}
v_{\textrm{Coul},i}^E(r)=-e^2\left ( \frac{3}{\pi} \right )^{1/3}\int d^3 \mathbf{r}^\prime \left[\rho_i(\mathbf{r}^\prime)\right]^{1/3}.
\end{equation}
where $i=\textrm{p},\Lambda_c$.
As mentioned above, the direct Coulomb potential is always repulsive while the exchange Coulomb potential is always attractive in the case of charmed hypernuclei.

\section{Results}\label{Results}

\subsection{The Ground State Properties of $\Lambda_c$ Hypernuclei}

In this section, starting from the lightest charmed hypernuclei \ce{^5_{$\Lambda_c$}Li}, we discuss the ground state properties of
\ce{^17_{$\Lambda_c$}F}, \ce{^41_{$\Lambda_c$}Sc}, \ce{^57_{$\Lambda_c$}Cu}, \ce{^133_{$\Lambda_c$}Sb} and \ce{^209_{$\Lambda_c$}Bi}.
%$^{17}$(O$+\Lambda_c$), $^{41}$(Ca$+\Lambda_c$), $^{57}$(Ni$+\Lambda_c$), ^{133}$(Sn$+\Lambda_c$) and $^{209}$(Pb$+\Lambda_c$).

\begin{figure}
\centering
\begin{subfigure}{0.49\textwidth}
\hspace{-0.25cm}
\includegraphics[width=1\textwidth]{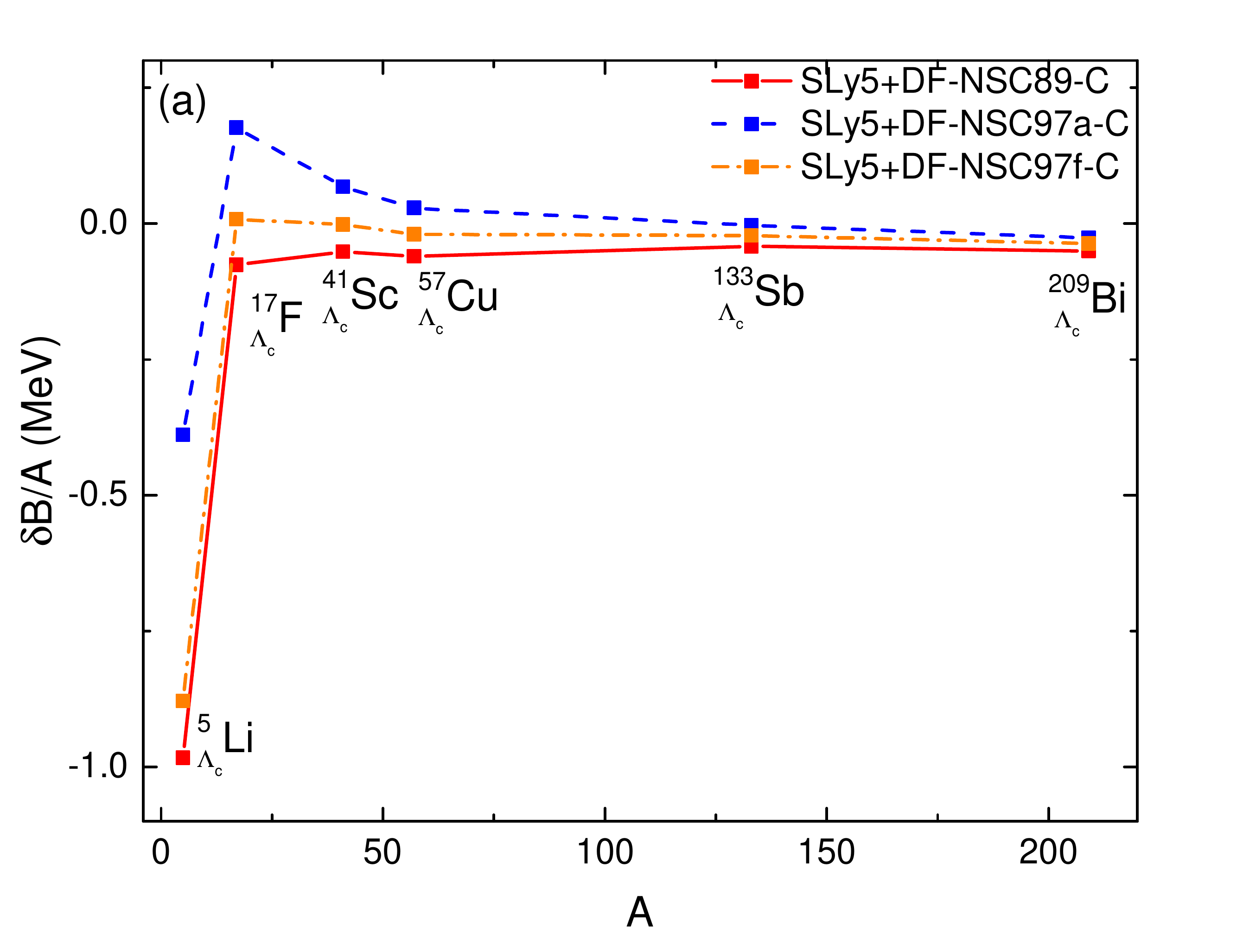}
\end{subfigure}
\begin{subfigure}{0.49\textwidth}
%\vspace{-1cm}
\includegraphics[width=1\textwidth]{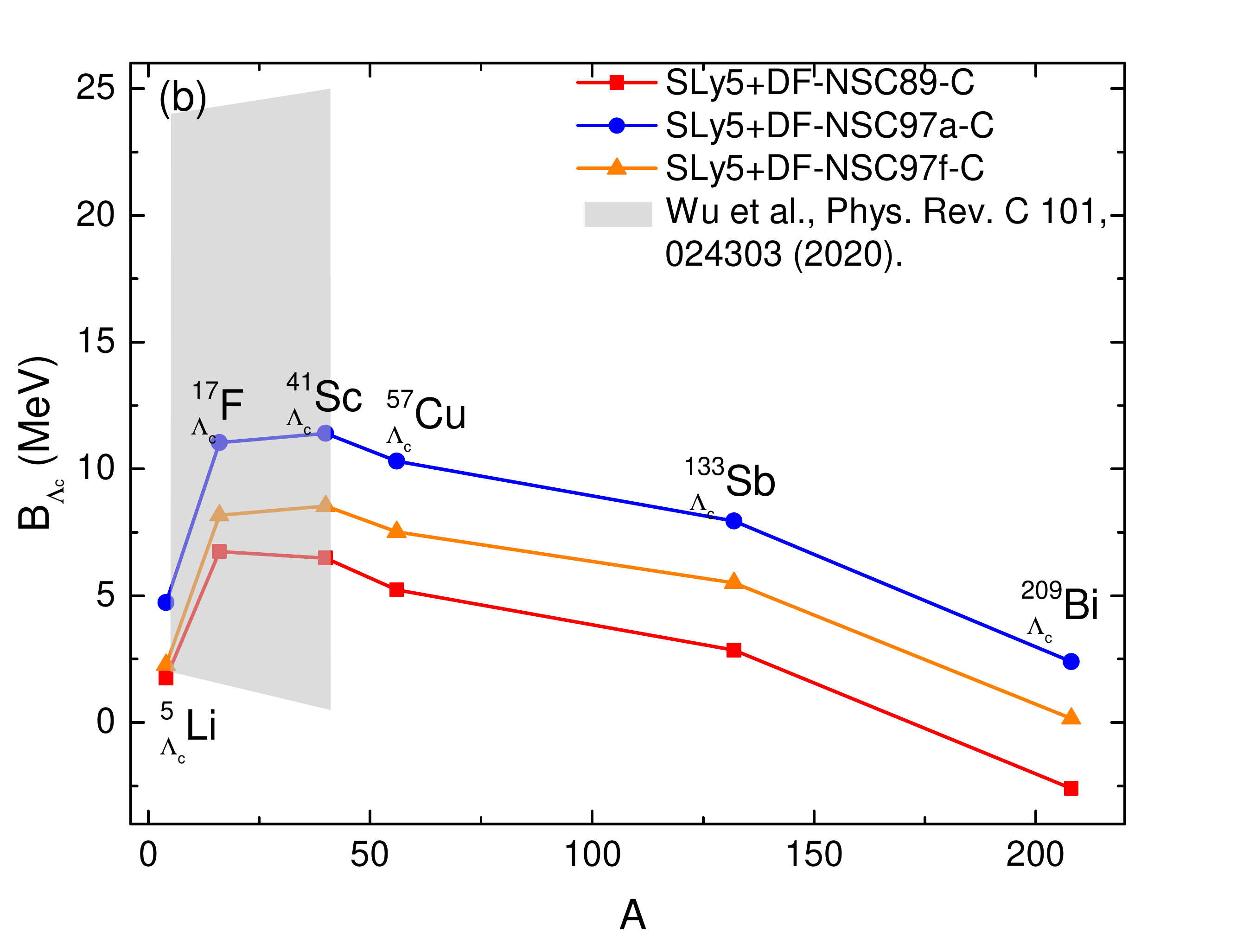}
\end{subfigure}
%\vspace{-1cm}
\caption{(a) The difference on binding energy per baryon $\delta B/A$, with respect to one of the nucleus without the charmed baryon, and (b) the $\Lambda_c$ separation energy $B_{\Lambda_c}$ for DF-NSC89-C, DF-NSC97a-C and DF-NSC97f-C force sets.}%\JM{Increase legend size}}
\label{f1}
%\vspace*{-3cm}
 \end{figure}

The difference on binding energy per baryon $\delta B/A=B/A(\ce{^{A+1}_{$\Lambda_c$}{Z+1}})-B/A(\ce{^{A}_{}Z})$, and the $\Lambda_c$ separation energy $B_{\Lambda_c}=E(\ce{^{A+1}_{$\Lambda_c$}{Z+1}})-E(\ce{^{A}_{}Z})$, are calculated for \ce{^5_{$\Lambda_c$}Li}, \ce{^17_{$\Lambda_c$}F}, \ce{^41_{$\Lambda_c$}Sc}, \ce{^57_{$\Lambda_c$}Cu}, \ce{^133_{$\Lambda_c$}Sb} and \ce{^209_{$\Lambda_c$}Bi} charmed hypernuclei, in Fig.~\ref{f1}.
%$\delta B/A= B/A \Big(^{A+1}_{\Lambda_c}(Z+1) \Big)-B/A \Big(^AZ \Big)$ and $B_{\Lambda_c}=E\Big(^{A+1}_{\Lambda_c}(Z+1) \Big)-E \Big(^AZ \Big)$ respectively.
Charmed nuclei between \ce{^17_{$\Lambda_c$}F} and \ce{^57_{$\Lambda_c$}Cu} are found more bound than their core nuclei, for DF-NSC97a-C, in contrast with the two other interactions. This is due to the more attractive nature of DF-NSC97a-C
(as the interactions depend on the densities, differences are $3$ MeV for 1s channel of \ce{^5_{$\Lambda_c$}Li} and $5$ MeV for 1s channel of \ce{^133_{$\Lambda_c$}Sb}).
%(changing with the densities, $3$ MeV for 1s channel of \ce{^5_{$\Lambda_c$}Li} and $5$ MeV for 1s channel of \ce{^133_{$\Lambda_c$}Sb}),
The binding energy differences for DF-NSC97f-C are of intermediate values, between DF-NSC97a-C and DF-NSC89-C. It also implies a slightly stronger bound of \ce{^17_{$\Lambda_c$}F} with respect to \ce{^16_{}O}, in the case of DF-NSC97f-C.

In the case of $\Lambda_c$ separation energies, shown in Fig.~\ref{f1}(b), the general behavior as a function of A is similar for all force sets, with a maximum difference of $6$ MeV between DF-NSC89-C and DF-NSC97a-C.
The maximum $\Lambda_c$ separation energies are predicted to occur between \ce{^17_{$\Lambda_c$}F} and \ce{^41_{$\Lambda_c$}Sc}.
Despite the decreasing trend for the A$\ge41$ region, $\Lambda_c$ separation energies indicate possible bound $\Lambda_c$ hypernuclei up to \ce{^209_{$\Lambda_c$}Bi}.
This is in contrast with Ref.~\cite{Miyamoto2018}, where its N-$\Lambda_c$ interaction (calculated from HAL QCD approach) predicts the existence of $\Lambda_c$ hypernuclei only up to A$=58$.
The results of Ref.~\cite{Wu2020} are also shown in Fig.~\ref{f1}(b) for comparison, where minimum and maximum limits are obtained from QMF-NK3C$^\prime$ and QMF-NK3C, respectively.
The main difference between QMF-NK3C$^\prime$ and QMF-NK3C is that the coupling constants of QMF-NK3C$^\prime$ are obtained from the HAL QCD collaboration, while the coupling constants of QMF-NK3C are obtained from the
naive quark counting rule, see Ref.~\cite{Wu2020}. Fig.~\ref{f1}(b) shows that our results are compatible with these predictions.

\begin{figure}
\centering
\begin{subfigure}{0.5\textwidth}
\hspace{-0.6cm}
\includegraphics[width=1\textwidth]{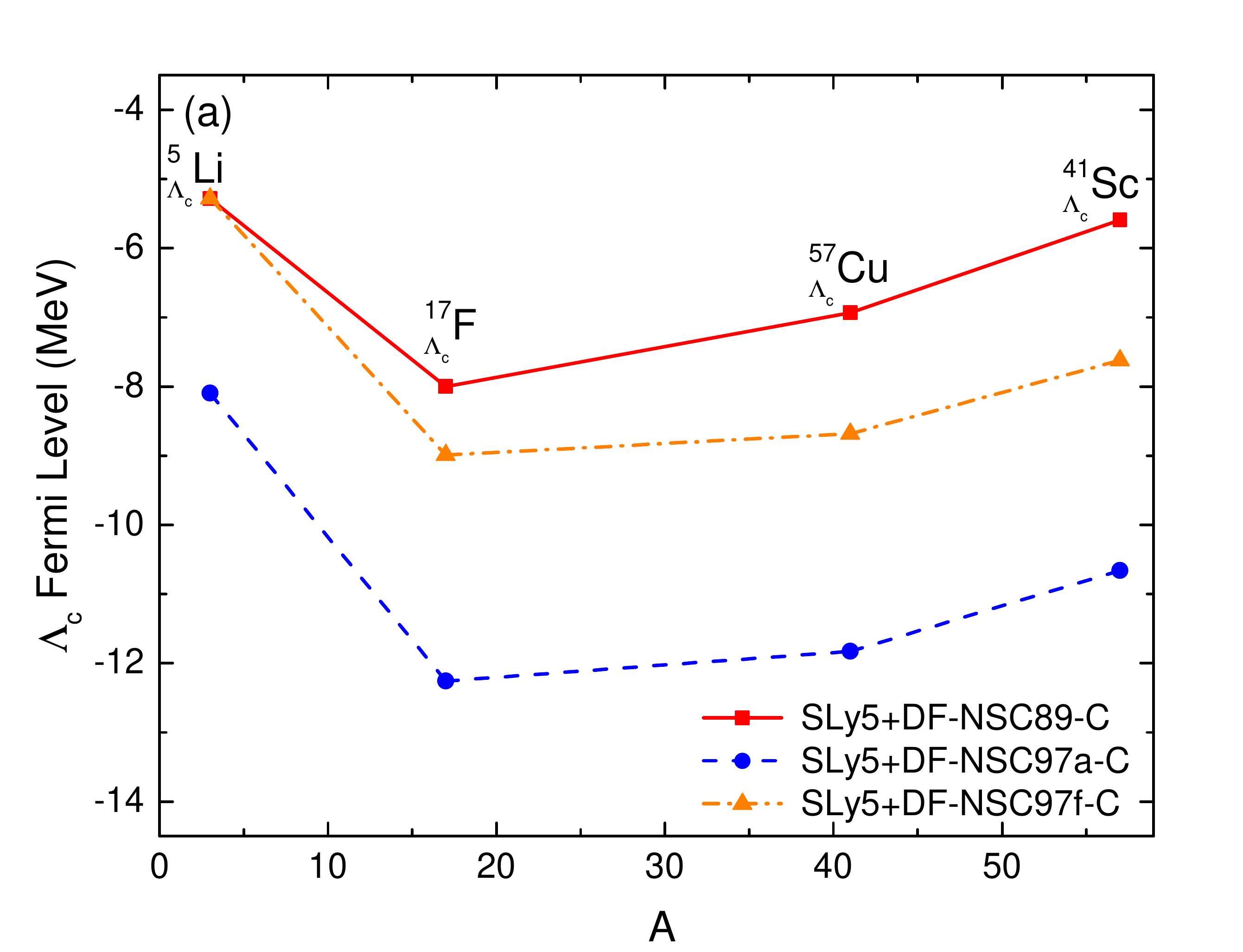}
\end{subfigure}
\begin{subfigure}{0.5\textwidth}
%\vspace{-1cm}
\includegraphics[width=1\textwidth]{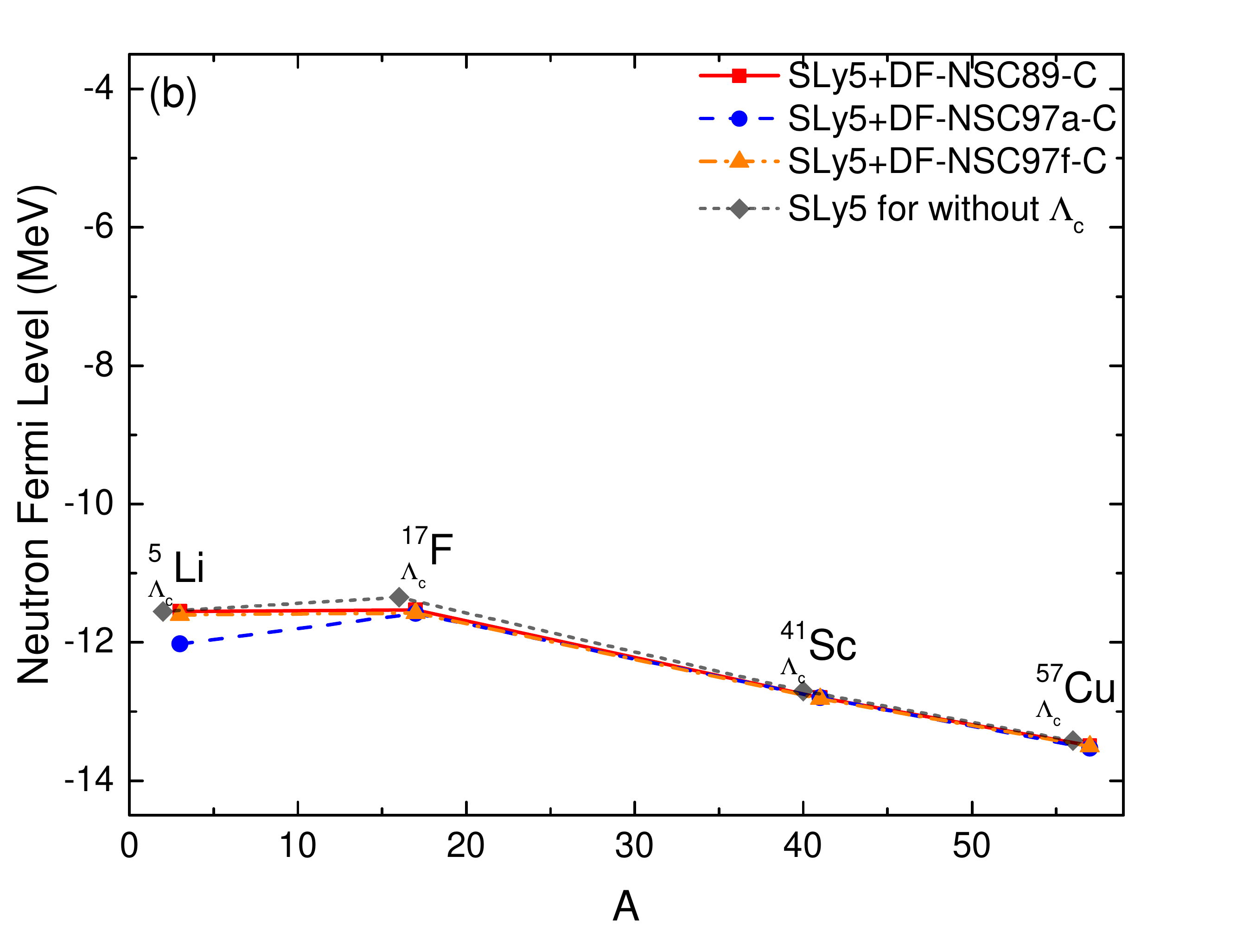}
\end{subfigure}
\begin{subfigure}{0.5\textwidth}
%\vspace{-1cm}
\includegraphics[width=1\textwidth]{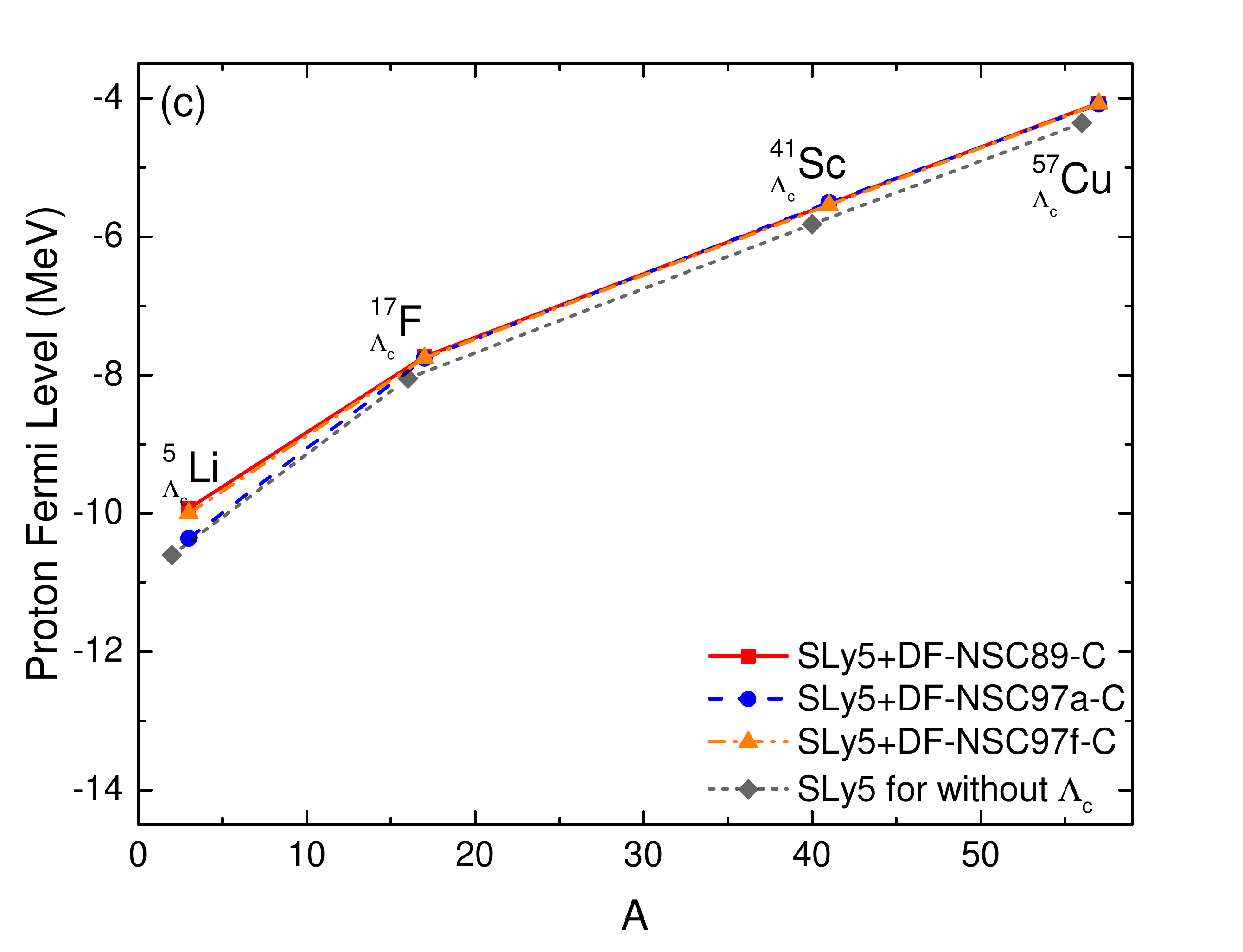}
\end{subfigure}
%\vspace{-1cm}
\caption{The Fermi energies of (a) $\Lambda_c$, (b) neutrons and (c) protons for DF-NSC89-C, DF-NSC97a-C and DF-NSC97f-C force sets.}
\label{f1.1}
%\vspace*{-3cm}
 \end{figure}

The $\Lambda_c$, neutron and proton Fermi levels are displayed in Fig.~\ref{f1.1}. As expected, small (order of keV) differences are observed on the proton and neutron Fermi levels for all force sets. However, there is an average $5$ MeV difference on $\Lambda_c$ Fermi energies between DF-NSC89-C and DF-NSC97a-C, and $2$ MeV difference between DF-NSC89-C and DF-NSC97f-C.
Due to the Coulomb repulsion, the most bound charmed hypernucleus is \ce{^17_{$\Lambda_c$}F}, where the energy of the 1s state is $-7$ MeV for DF-NSC89-C, $-12$ MeV for DF-NSC97a-C and $-8$ MeV for DF-NSC97f-C.
Since \ce{^17_{$\Lambda_c$}F} is predicted with the lowest value for the binding energy, an unambiguous signature of the existence of charmed hypernucleus may be found by producing this system.
%Therefore, \ce{^17_{$\Lambda_c$}F} could be an excellent candidate for measuring charmed hypernucleus.
For charmed nuclei heavier than \ce{^17_{$\Lambda_c$}F}, the binding energy decreases, resulting in less bound $\Lambda_c$ states, due to the Coulomb repulsion.
%For charmed nuclei heavier than \ce{^17_{$\Lambda_c$}F}, the binding energy decreases and results in unbound state for \ce{^209_{$\Lambda_c$}Bi} charmed hypernucleus for DF-NSC89-C and DF-NSC97f-C.
%However, DF-NSC97a-C has a strong enough attraction to counter balance the Coulomb repulsion leading to bound \ce{^209_{$\Lambda_c$}Bi} charmed hypernucleus.

\begin{figure}
\centering
\begin{subfigure}{0.5\textwidth}
\includegraphics[width=1\textwidth]{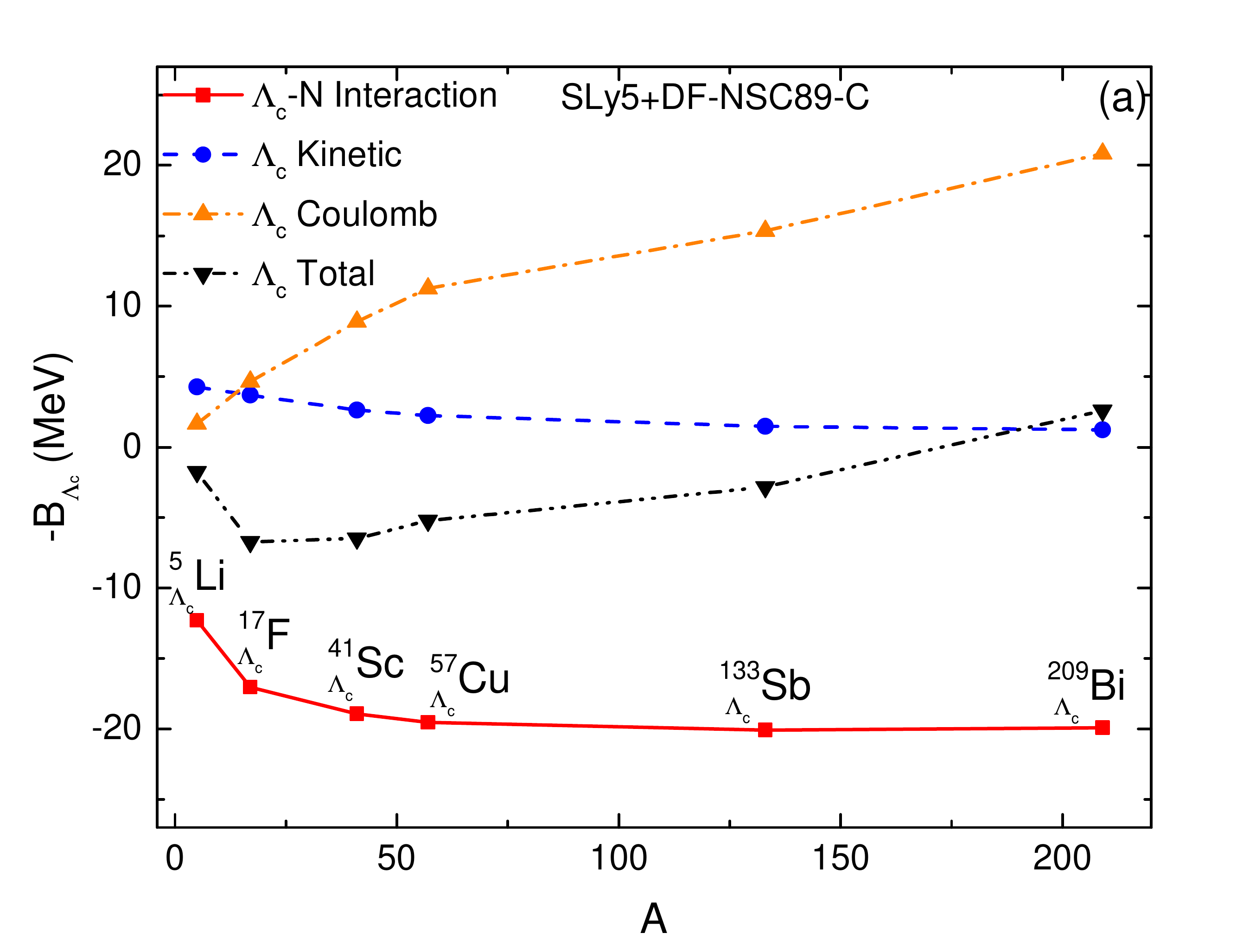}
\end{subfigure}
\begin{subfigure}{0.5\textwidth}
%\vspace{-1cm}
\includegraphics[width=1\textwidth]{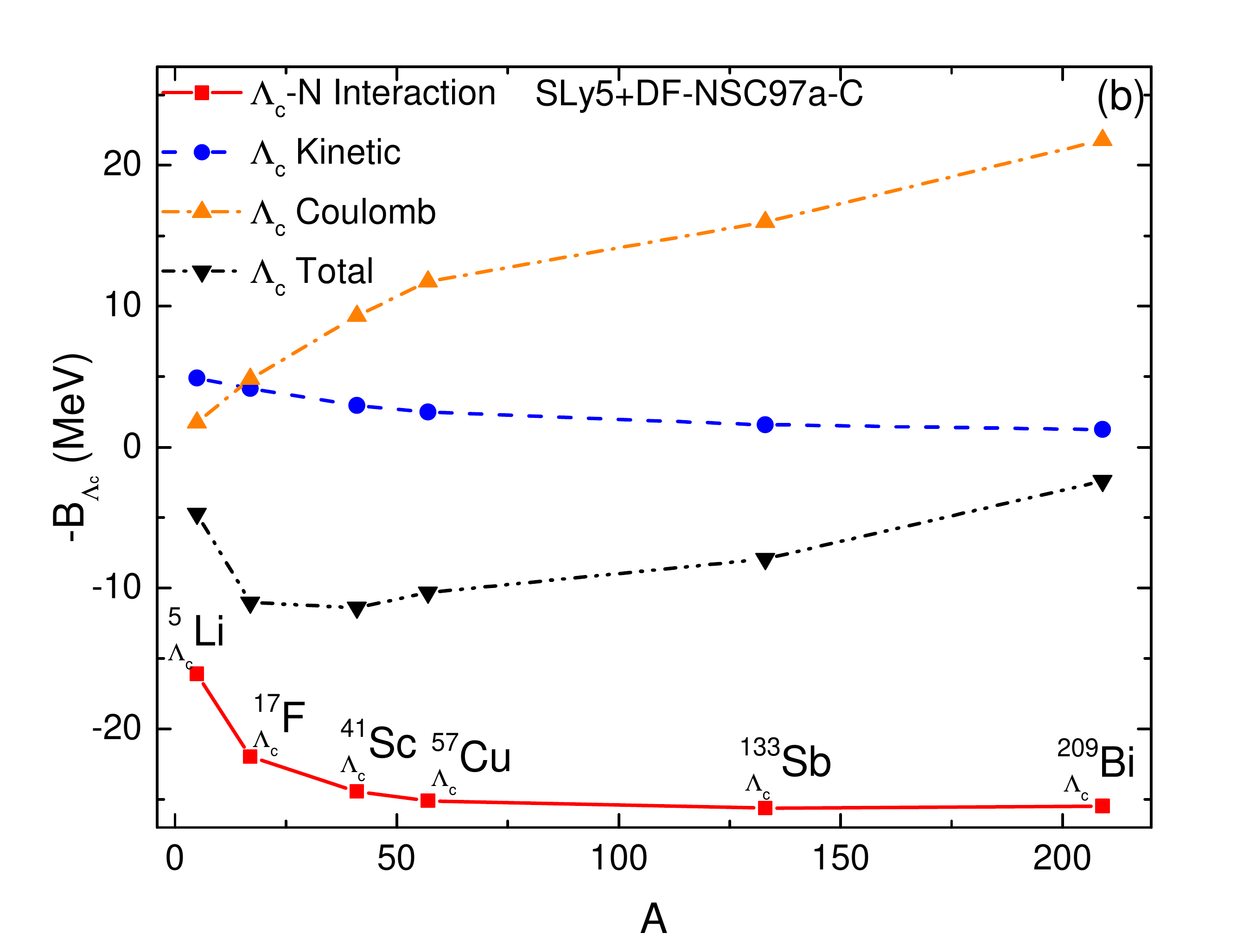}
\end{subfigure}
\begin{subfigure}{0.5\textwidth}
%\vspace{-1cm}
\includegraphics[width=1\textwidth]{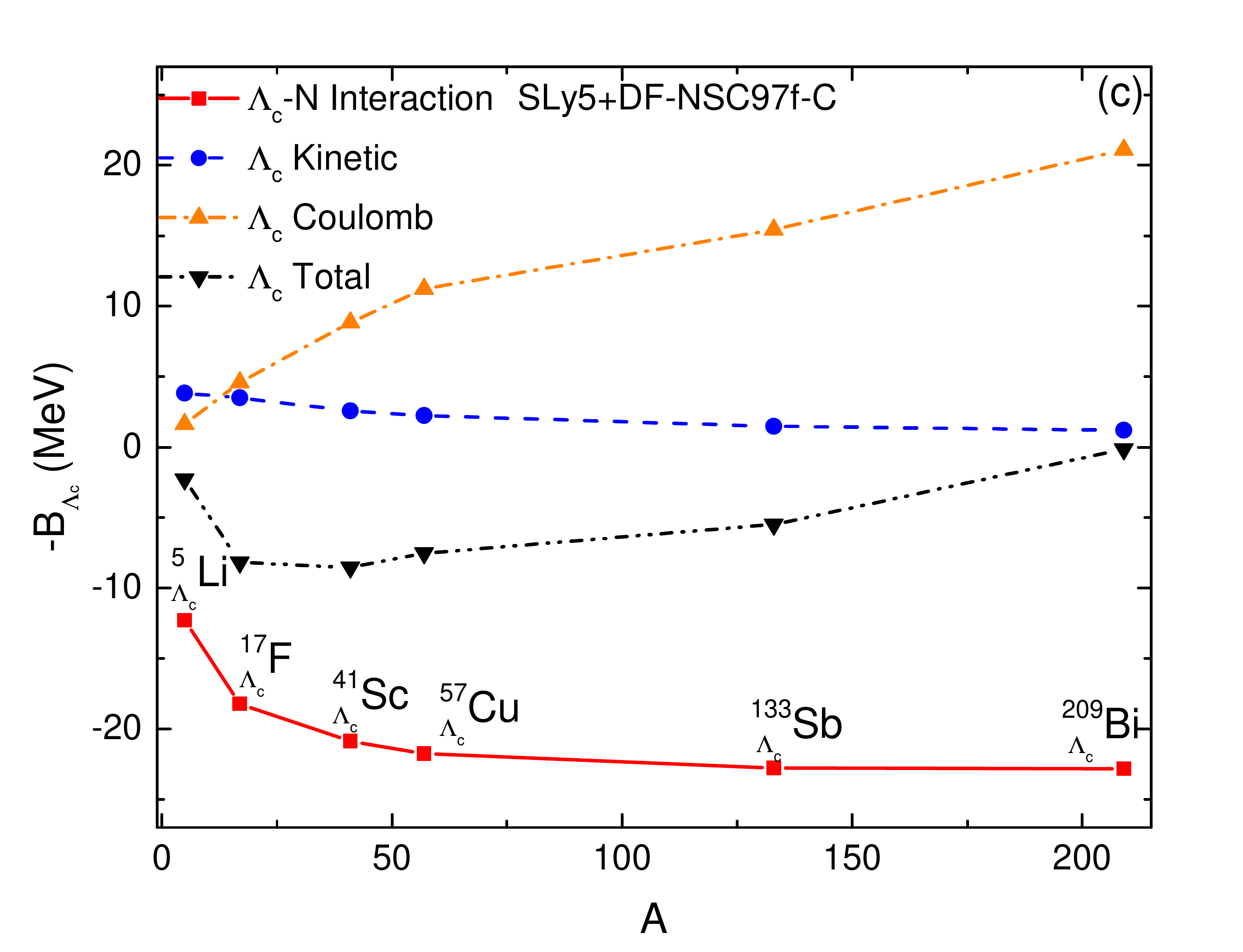}
\end{subfigure}
%\vspace{-1cm}
\caption{Contributions of the kinetic energy, N-$\Lambda_c$ interaction and the Coulomb potential on the $\Lambda_c$ separation energy for (a) DF-NSC89-C, (b) DF-NSC97a-C and (c) DF-NSC97f-C force sets.}
\label{s2}
%\vspace*{-3cm}
 \end{figure}

\begin{figure}
\centering
\begin{subfigure}{0.5\textwidth}
\includegraphics[width=1\textwidth]{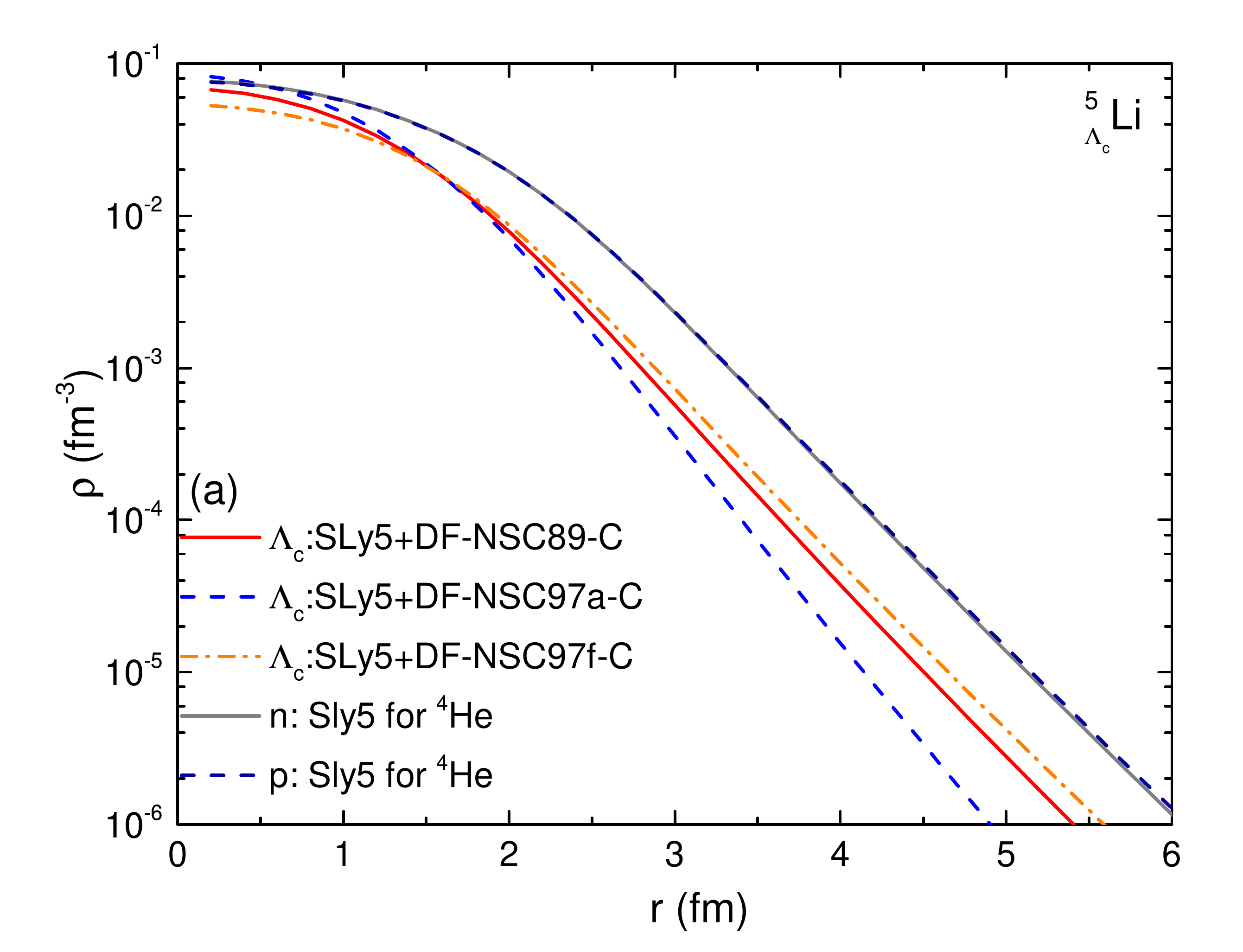}
\end{subfigure}
\begin{subfigure}{0.5\textwidth}
%\vspace{-1cm}
\includegraphics[width=1\textwidth]{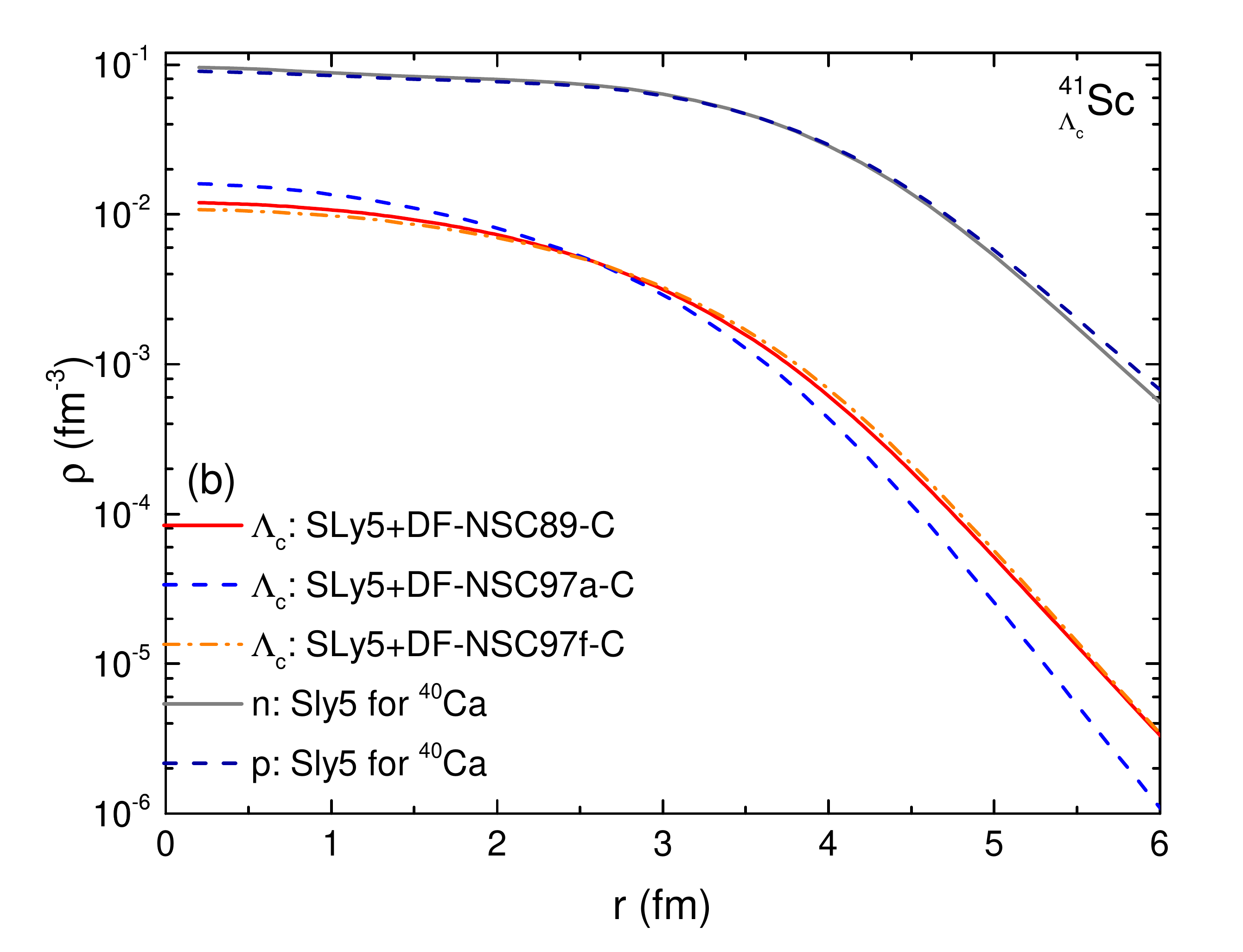}
\end{subfigure}
\begin{subfigure}{0.5\textwidth}
\hspace{-0.9cm}
\includegraphics[width=1\textwidth]{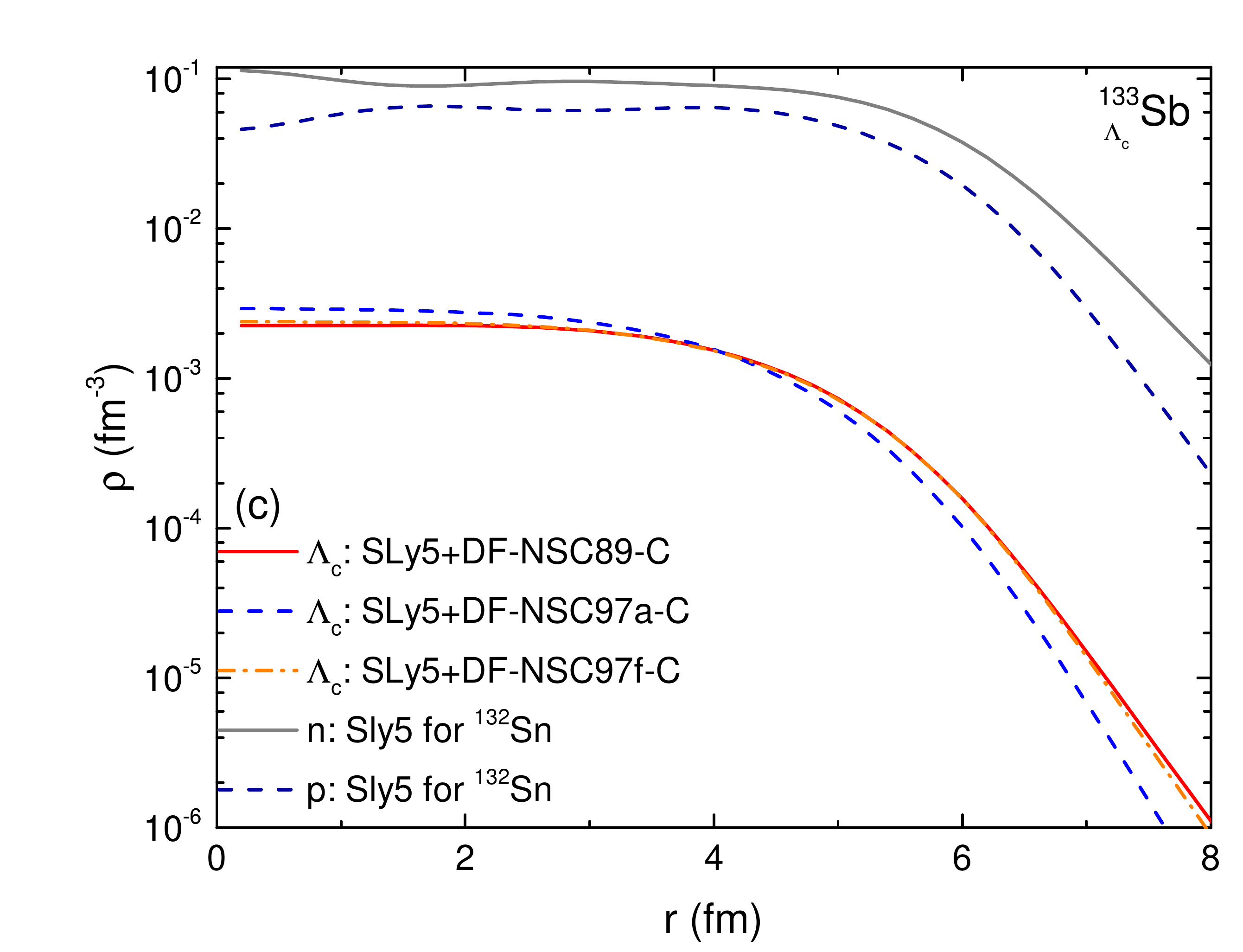}
\end{subfigure}
\caption{The $\Lambda_c$ density distribution of (a) \ce{^5_{$\Lambda_c$}Li}, (b) \ce{^41_{$\Lambda_c$}Sc} and (c)  \ce{^133_{$\Lambda_c$}Sb} for DF-NSC89-C, DF-NSC97a-C and DF-NSC97f-C force sets.}
\label{s3}
%\vspace*{-3cm}
\end{figure}

In order to investigate in more details this behavior, the contributions of the kinetic energy, N-$\Lambda_c$ interaction and the Coulomb potential, to the $\Lambda_c$ separation energy are shown in Fig.~\ref{s2}.
%$\epsilon_{\Lambda_c}=-B_{\Lambda_c}$ to emphasize attractive and repulsive nature of interactions.
For all charmed nuclei, DF-NSC97a-C predicts a more bound system, compared to other interactions ($\Lambda_c$ Total on Fig.~\ref{s2}).
The Coulomb repulsion has a increasing behavior proportional to the mass number,
while the N-$\Lambda_c$ interaction saturates around \ce{^57_{$\Lambda_c$}Cu} for all force sets.
These effects explain the smaller separation energies in heavier charmed nuclei than \ce{^41_{$\Lambda_c$}Sc}.
The saturation property of N-$\Lambda_c$ interactions is compatible with the results of HAL QCD for the A$>50$ region~\cite{Miyamoto2018}.
However, the present N-$\Lambda_c$ interactions are attractive enough to create bound $\Lambda_c$ hypernuclei in the heavy mass region.
%Due to both increasing Coulomb repulsion and the saturation behavior of N-$\Lambda_c$,
Due to the saturation behavior of N-$\Lambda_c$,
Ref.~\cite{Miyamoto2018} suggests that only light or medium-mass charmed hypernuclei could really exist.
%This is also in contrast with recent work of Vida\~{n}a et al., (See Ref.~\cite{Vidana2019} for details) where stronger N-$\Lambda_c$ attraction were found, with increasing mass number.
This is also in contrast with the work of Vida\~{n}a et al. and Haidenbauer et al. (see Refs.~\cite{Vidana2019,Haidenbauer2020} for details) where stronger N-$\Lambda_c$ attraction were found, as in our case.

In order to investigate the spatial properties of charmed hypernuclei, $\Lambda_c$ density distributions are displayed in Fig.~\ref{s3}, together with the proton and neutron distributions of the core nuclei.
For all charmed hypernuclei, DF-NSC97a-C generates more packed distributions in the centre of the nucleus, especially for the lightest case (\ce{^5_{$\Lambda_c$}Li}).
Due to the Coulomb repulsion, the $\Lambda_c$ distribution extends to larger radii, in heavier hypernuclei. This dilution of the charmed baryon, in the nucleus, impacts its density at the centre, which can decrease by a couple of orders of magnitude, from the lightest to the heaviest charmed hypernuclei.

\subsection{Comparison of $\Lambda_c$ vs. $\Lambda$ in Hypernuclei}

It could be relevant to compare the properties of hypernuclei and charmed ones, to understand for instance the respective role of the Coulomb interaction as well as of the scaling $N\Lambda_c$ interaction factor $K$. On this purpose, $\Lambda$ hypernuclei are calculated with the same HF scheme, using DF-NSC89, DF-NS97a and DF-NSC97f force sets, as introduced in Table~\ref{t1}.
Starting from \ce{^5_{$\Lambda$}He}, we compare \ce{^17_{$\Lambda$}O}, \ce{^41_{$\Lambda$}Ca}, \ce{^57_{$\Lambda$}Ni}, \ce{^133_{$\Lambda$}Sn} and \ce{^209_{$\Lambda$}Pb} hypernuclei with their charmed counterparts.

\begin{figure}
  \centering
  \begin{subfigure}{0.5\textwidth}
  \hspace{-0.9cm}
  \includegraphics[width=1\textwidth]{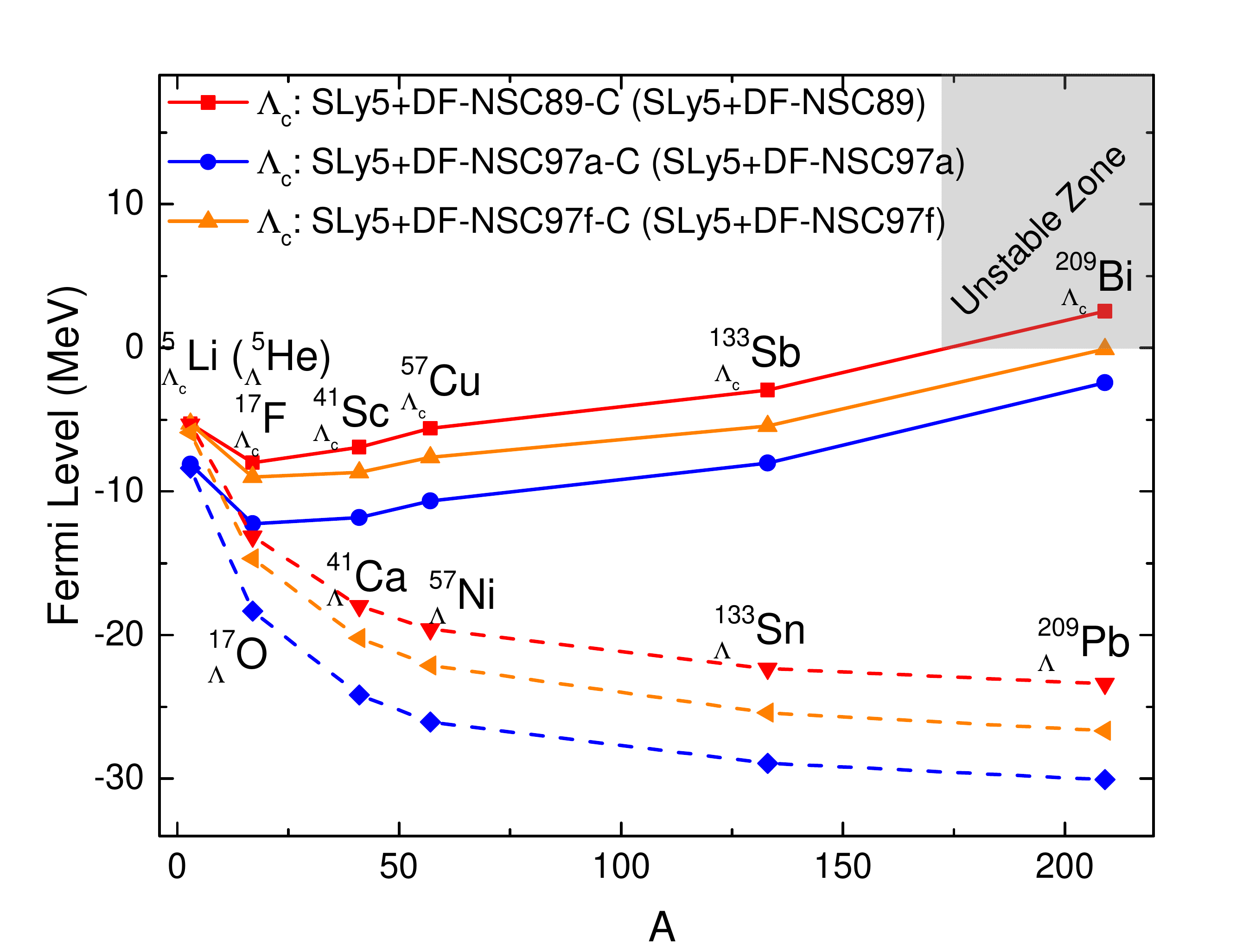}
  \end{subfigure}
  \caption{The Fermi energy of the $\Lambda_c$ ($\Lambda$) for the DF-NSC89-C (DF-NSC89), DF-NSC97a-C (DF-NSC97a) and DF-NSC97f-C (DF-NSC97f) force sets, in solid lines (dashed lines).
  Unbound hypernuclei are displayed in the gray area (Unstable Zone).}
  \label{s4}
\end{figure}

\begin{figure}
\centering
\begin{subfigure}{0.5\textwidth}
\includegraphics[width=1\textwidth]{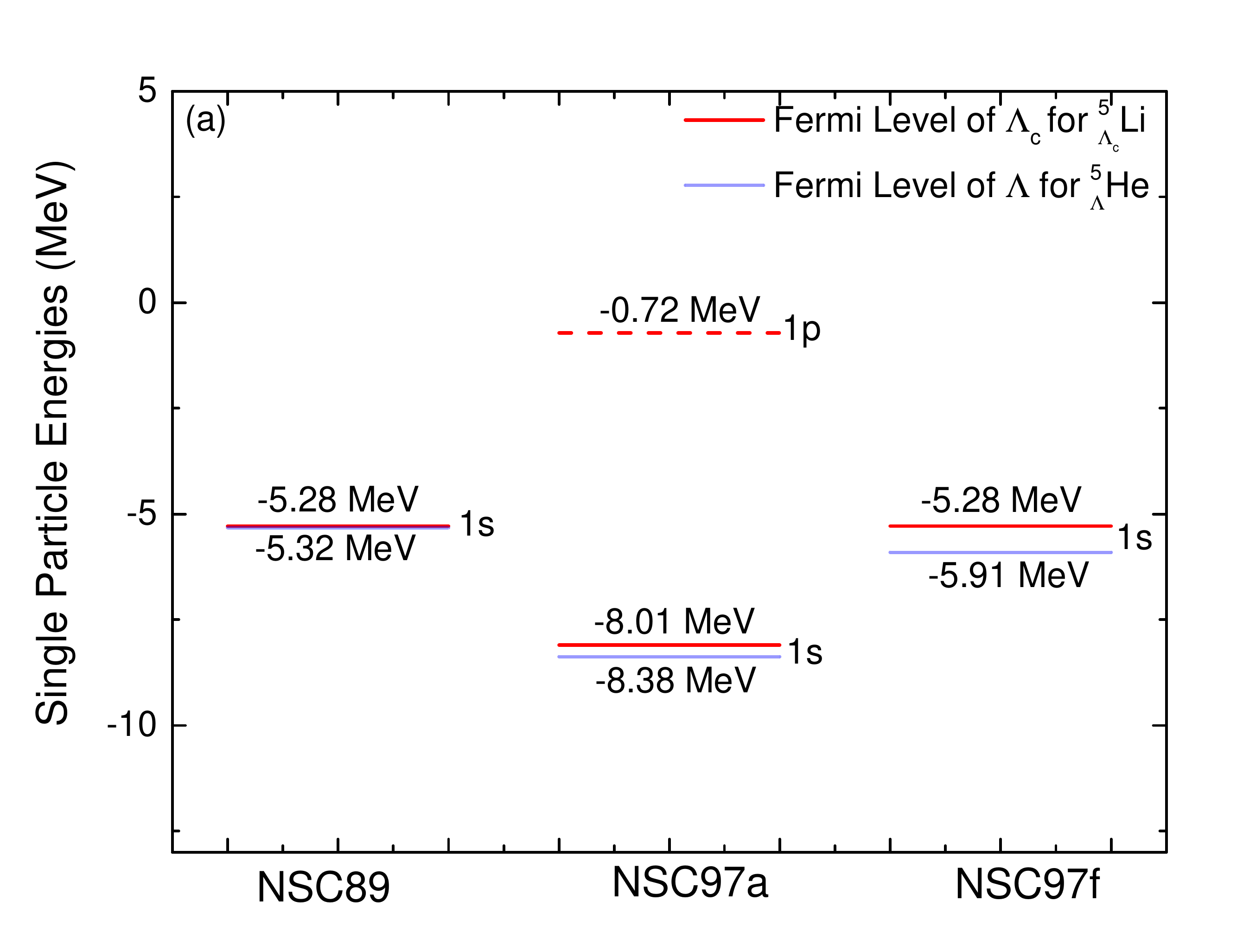}
\end{subfigure}
\begin{subfigure}{0.5\textwidth}
%\vspace{-1cm}
\includegraphics[width=1\textwidth]{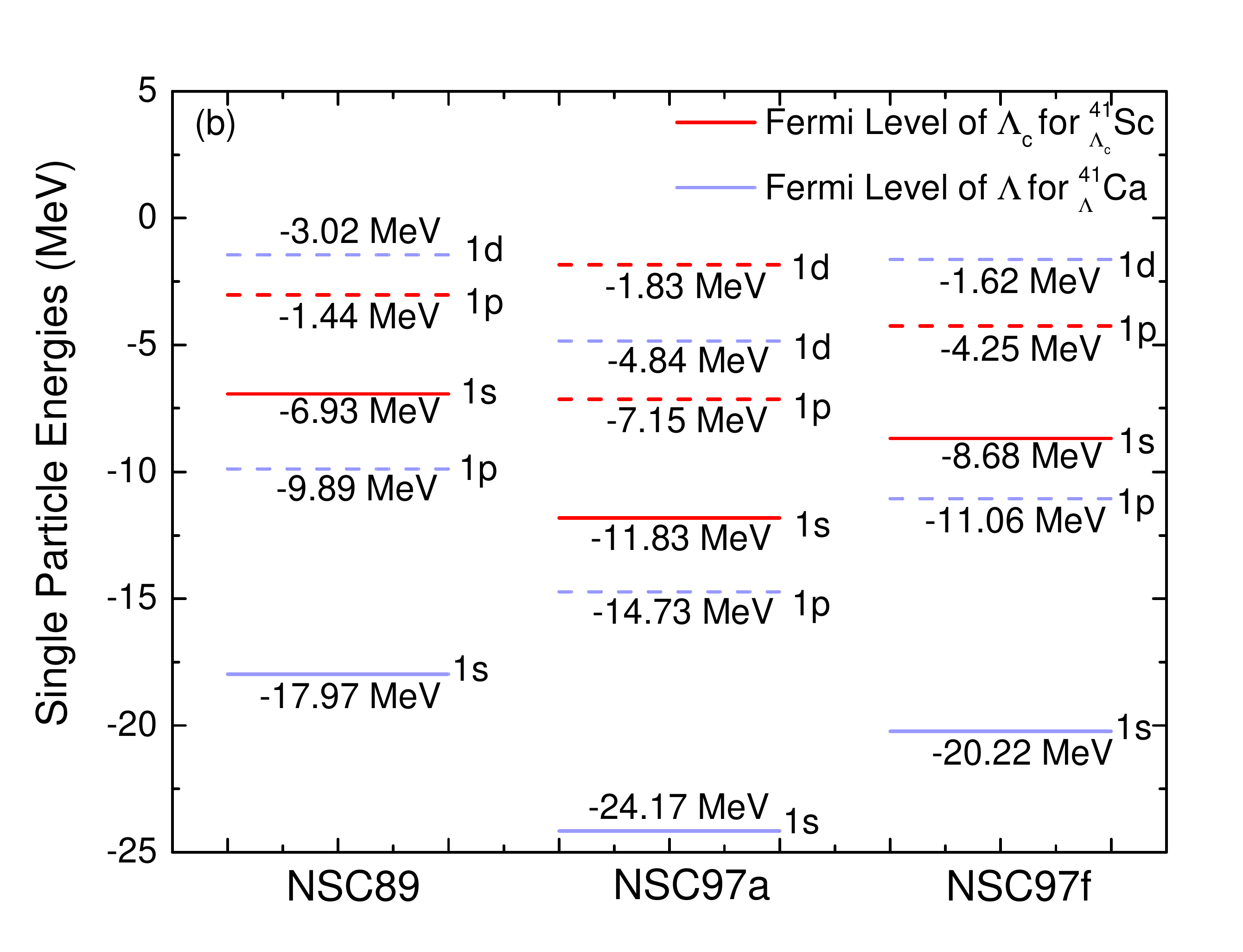}
\end{subfigure}
\begin{subfigure}{0.5\textwidth}
%\vspace{-1cm}
\includegraphics[width=1\textwidth]{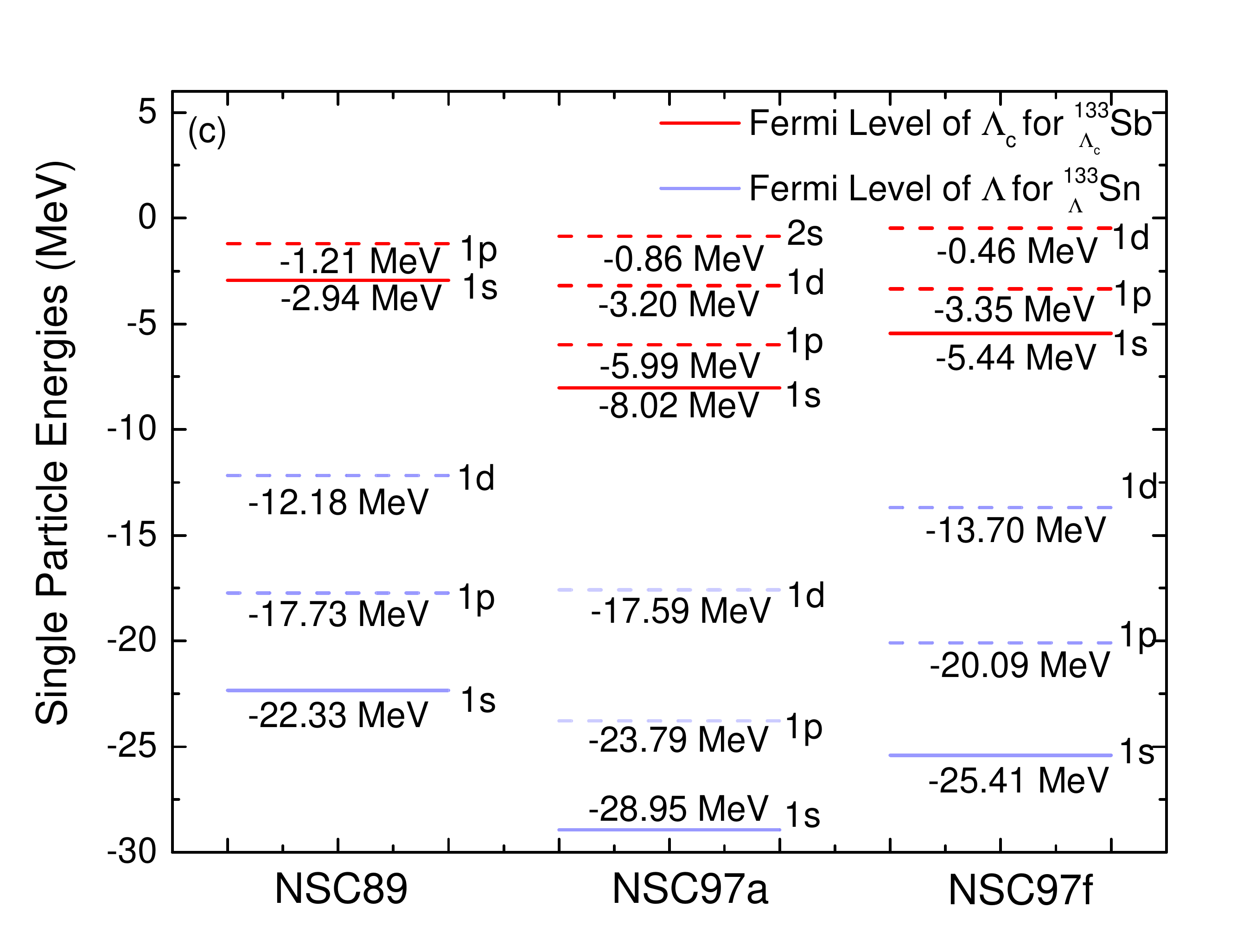}
\end{subfigure}
%\vspace{-1cm}
\caption{The $\Lambda_c$ ($\Lambda$) single particle spectrum of \ce{^5_{$\Lambda_c$}Li} (\ce{^5_{$\Lambda$}He}) (a), \ce{^41_{$\Lambda$}Sc} (\ce{^41_{$\Lambda$}Ca}) (b)
and \ce{^133_{$\Lambda$}Sb} (\ce{^133_{$\Lambda$}Sn}) (c) for NSC89, NSC97a and NSC97f force sets.}
\label{s5}
%\vspace*{-3cm}
 \end{figure}

We first represent Fermi energies for $\Lambda_c$ and $\Lambda$ in Fig.~\ref{s4}.
In the case of the lightest \ce{^5_{$\Lambda_c$}Li}-\ce{^5_{$\Lambda$}He} pair, Fermi energies are nearly identical, between the charmed and normal NSC force sets.
However, both $\Lambda$ and $\Lambda_c$ Fermi levels increases (in absolute value), up to the \ce{^17_{$\Lambda$}O}-\ce{^17_{$\Lambda_c$}Li} pair, where
the Fermi energy differences between $\Lambda$ and $\Lambda_c$ are $5.02$ MeV for NSC89, $5.92$ MeV for NSC97a and $5.81$ MeV for NSC97f.
But, as the mass number increases, larger energy differences occur, due to the Coulomb repulsion, especially after the A$=17$ region. This  leads to unbound state for \ce{^209_{$\Lambda_c$}Bi} charmed hypernucleus for DF-NSC89-C and DF-NSC97f-C. The increasing effect of the Coulomb repulsion is also the main reason
why \ce{^17_{$\Lambda_c$}F} is an excellent test nucleus, to probe charmed hypernuclei: the effect of the N-$\Lambda_c$ interaction is expected to be noticeable and therefore could be probed in order to constrain it.
It should be noted that DF-NSC97a-C has a large enough attraction, to counter balance the Coulomb repulsion, leading to bound \ce{^209_{$\Lambda_c$}Bi} charmed hypernucleus. Also, the maximum Fermi energy differences are spotted at \ce{^209_{$\Lambda$}Pb}-\ce{^209_{$\Lambda_c$}Bi} pair: $26$ MeV for NSC89 (and also for NSC97a), and $32$ MeV for NSC97f.

Finally, the $\Lambda_c$ ($\Lambda$) single particle energies are displayed in Fig.~\ref{s5}:
for a light charmed hypernucleus \ce{^5_{$\Lambda_c$}Li} (\ce{^5_{$\Lambda$}He}), for a medium-mass charmed hypernucleus \ce{^41_{$\Lambda$}Sc} (\ce{^41_{$\Lambda$}Ca}) and
for heavy charmed hypernucleus \ce{^133_{$\Lambda$}Sb} (\ce{^133_{$\Lambda$}Sn}).
In the \ce{^5_{$\Lambda_c$}Li} case, DF-NSC89-C and DF-NSC97f-C predict similar values for the 1s state.
However, DF-NSC97a-C differs by $2.72$ MeV from DF-NSC89-C and DF-NSC97f-C.
Since the Coulomb repulsion is rather weak for the case of \ce{^5_{$\Lambda_c$}Li},
similar results are obtained in the case of hypernuclei: in \ce{^5_{$\Lambda$}He},
there is $0.59$ MeV difference between DF-NSC89 and DF-NSC97f,
but $3.10$ MeV between DF-NSC89 and DF-NSC97a.
In addition, DF-NSC89-C and DF-NSC97f-C do not allow for any excited state in the \ce{^5_{$\Lambda_c$}Li} case, whereas
DF-NSC89a-C allows for such a state.

The situation starts to change for the \ce{^41_{$\Lambda$}Sc} case, where DF-NSC89-C and DF-NSC97f-C differ from each other, by $1.76$ MeV for the 1s state.
The gap, between 1s and 1p states, is $3.93$ MeV for DF-NSC89-C, $4.43$ MeV for DF-NSC97f-C and $4.68$ MeV for DF-NSC97a-C.
DF-NSC89-C and DF-NSC97f-C allow only 1p state while DF-NSC97a-C also allows for a 1d state in addition. Therefore, a possible spectroscopy of charmed hypernuclei, could allow to disentangle between the various interations.
In the case of \ce{^41_{$\Lambda$}Ca}, 1s state energy levels more bound because of the Coulomb repulsion:
such energy differences on the 1s state are $11.04$ MeV for NSC89, $17.02$ MeV for NSC-97a and $11.54$ MeV for NSC-97f.
In the \ce{^133_{$\Lambda$}Sb} case, the predicted levels are different for each force set: charmed nuclei are becoming less bound due to the Coulomb repulsion. An average of $20$ MeV difference is predicted for the 1s state between \ce{^133_{$\Lambda$}Sn} and \ce{^133_{$\Lambda$}Sb}.
These results are in agreement with those previously obtained by RMF and QMC models \cite{Tan2004,Wu2020}.
More precisely, the present $\Lambda_c$ 1s state predictions are located between those of  QMF-NK3C and QMF-NK3C$^\prime$ ~\cite{Wu2020}, for all charmed hypernuclei.

\section{Conclusions}~\label{Conclusions}

In this work we have investigated the ground state properties of charmed hypernuclei, close to doubly magic closed-shell nuclei.
We use SLy5 Skyrme force for the NN interaction and we generate N$\Lambda_c$ interactions from the N$\Lambda$ interactions obtained from microscopic Brueckner-Hartree-Fock calculations.

Since $\Lambda_c$ is positively charged, the Coulomb interaction plays an important role in the ground state properties of charmed hypernuclei.
Additionally, it is shown that N$\Lambda_c$ interactions have a saturation property around \ce{^57_{$\Lambda_c$}Cu}.
Due to these reasons, $\Lambda_c$ is maximally bound for mass numbers between \ce{^17_{$\Lambda_c$}F} and \ce{^41_{$\Lambda_c$}Sc}.
\ce{^17_{$\Lambda_c$}F} is an excellent candidate for the experimental measurements of charmed hypernuclei, since the predictions from the different functionals considered in this study are the largest one and the most different for this system.
For charmed hypernuclei heavier than \ce{^41_{$\Lambda_c$}Sc}, Coulomb repulsion becomes comparable in absolute value with the attractive N$\Lambda_c$ interactions, and charmed hypernuclei become less and less bound as $A$ increases, leading to unbound \ce{^209_{$\Lambda_c$}Bi} (for DF-NSC89-C and DF-NSC97f-C interactions).
Our results confirm previous ones obtained by the HAL QCD collaboration: \ce{^209_{$\Lambda_c$}Bi} is unbound and the N$\Lambda_c$ interaction saturates in the mid-$A$ region.
However, DF-NSC97a-C has a strong enough attraction to counter balance the Coulomb repulsion, leading to weakly-bound \ce{^209_{$\Lambda_c$}Bi} charmed hypernucleus.

In addition, the gap between 1s and 1d states is found to be about $4$ MeV for the majority of charmed hypernuclei. As a final conclusion, the reason of the uncertainties in the N$\Lambda_c$ interaction is mostly due to the lack of experimental measurements. Our results, confronted to others, predict different behavior for charmed hypernuclei as function of the mass number $A$, for which the future experimental facilities will help in selecting among the best density functional approaches.

\section{Acknowledgment}

This work is supported by the Scientific and Technological Research Council of Turkey (T\"{U}B\.{I}TAK) under project number MFAG-118F098 and the Yildiz Technical University under project number FBI-2018-3325 and FBA-2021-4229.
E.K. and J.M. are both supported by the CNRS/IN2P3 NewMAC project, and are also grateful to PHAROS COST Action MP16214. J.M. is grateful to the LABEX Lyon Institute of Origins (ANR-10-LABX-0066) of the \textsl{Universit\'e de Lyon} for its financial support within the program \textsl{Investissements d'Avenir} (ANR-11-IDEX-0007) of the French government operated by the National Research Agency (ANR).

% >>>>>>>>>>>>>>>>>>>>>>>>>>>>>>>>>>>>>>>>>>>>>>>>>>>>>>>>>>>>>>>>>>>>
% REFERENCES.
%\clearpage

%

% >>>>>>>>>>>>>>>>>>>>>>>>>>>>>>>>>>>>>>>>>>>>>>>>>>>>>>>>>>>>>>>>>>>>
\end{document}